\documentclass[manuscript,screen]{acmart}

\usepackage{rotating}
\usepackage{adjustbox}
\usepackage{pgf}
\usepackage{tikz}
\usepackage{tcolorbox}
\usepackage{soul}
\usepackage{pdflscape}
\usepackage{multirow}
\usepackage{multicol}
\usepackage{blindtext}
\usepackage{tabularx}
\usepackage{graphicx}
\usepackage{subcaption}
\usepackage{dcolumn}
\usepackage{rotating}
\usepackage{xspace}
\usepackage[commandnameprefix=always]{changes}
\usepackage{changes}  
\definecolor{lightblue}{HTML}{dfe7fd}
\setaddedmarkup{\textcolor{blue}{#1}}
\colorlet{Changes@Color}{blue}
\AtBeginDocument{%
  }

\setcopyright{acmlicensed}
\copyrightyear{2018}
\acmYear{2018}
\acmDOI{XXXXXXX.XXXXXXX}
\acmConference[Conference acronym 'XX]{Make sure to enter the correct
  conference title from your rights confirmation email}{June 03--05,
  2018}{Woodstock, NY}
\acmISBN{978-1-4503-XXXX-X/2018/06}

\begin{document}

\title{Passing the Buck to AI: How Individuals' Decision-Making Patterns Affect Reliance on AI}

\author{Katelyn Xiaoying Mei}
\affiliation{%
  \institution{Information School, University of Washington}
     \country{USA}
}
\email{kmei@uw.edu}

\author{Rock Yuren Pang}
\affiliation{%
  \institution{Paul G. Allen School of Computer Science, University of Washington}
   \country{USA}
}
\email{ypang2@cs.washington.edu}

\author{Alex Lyford}
\affiliation{%
  \institution{Middlebury College}
   \country{USA}
}
\email{alyford@middlebury.edu}

\author{Lucy Lu Wang}
\affiliation{%
  \institution{Information School, University of Washington}
  \country{USA}
}
\email{lucylw@uw.edu}

\author{Katharina Reinecke}
\affiliation{%
  \institution{Paul G. Allen School of Computer Science,
  University of Washington}
 \country{USA}
}
\email{reinecke@cs.washington.edu}

\renewcommand{\shortauthors}{Mei et al.}

\newcommand\lucy[1]{{\color{blue}\{\textit{#1}\}$_{lucy}$}}
\newcommand\kr[1]{{\color{red}\{\textit{#1}\}$_{kr}$}}
\newcommand{\rp}[1]{\textcolor{green}{#1}}

\newcommand\todoit[1]{{\color{red}\{TODO: \textit{#1}\}}}

\newcommand\todocite{{\color{red}{CITE}}\xspace}

\definecolor{labbg}{RGB}{255, 217, 51}
\newtcbox{\ilabel}[1][]{
 box align=base,
 nobeforeafter,
 colback=labbg,
 colframe=labbg,
 size=small,
 fontupper=\scriptsize\bf\sffamily,
 left=0.5pt,
 right=0.5pt,
 top=0.5pt,
 bottom=0.5pt,
 boxsep=2pt,
 arc=4.5pt,
 #1}

\definecolor{lgreen}{RGB}{160, 217, 53} 

\newtcbox{\plabel}[1][]{
  nobeforeafter,
  colback=lgreen,
  colframe=lgreen,
  box align=base,
  size=small,
  fontupper=\scriptsize\bf\sffamily,
  left=0.5pt,
  right=0.5pt,
  top=0.5pt,
  bottom=0.5pt,
  boxsep=2pt,
  arc=0pt, 
  #1}


\newcommand{\hlc}[2][yellow]{%
    \begingroup
    \sethlcolor{#1}%
    \hl{#2}%
    \endgroup
}

\definecolor{green}{HTML}{94d2bd} 
\definecolor{turquoise}{HTML}{6CC9D6}
\definecolor{purple}{HTML}{D3A9F4}
\definecolor{orange}{HTML}{F9D78D}
\definecolor{yellow}{HTML}{f9c74f}

\definecolor{fallred}{HTML}{F6A877}






\newcommand{\seeAIdecision}{
    \hlc[green]{see AI decision}\xspace
}

\newcommand{\seeingAIdecision}{%
    \hlc[green]{Seeing AI Decision}\xspace
}

\newcommand{\seeAIexplanations}{%
    \hlc[yellow]{seeing AI explanations}\xspace
}

\newcommand{\toseeAIexplanations}{
    \hlc[yellow]{see AI explanations}\xspace
}

\newcommand{\seeingAIexplanations}{%
    \hlc[yellow]{Seeing AI Explanations}\xspace
}

\newcommand{\perceivedAIreliance}{%
    \hlc[fallred]{reliance on AI}\xspace
}

\newcommand{\PerceivedAIreliance}{%
    \hlc[fallred]{Perceived AI Reliance}\xspace
}

\begin{abstract}
Psychological research has identified different patterns individuals have while making decisions, such as vigilance (making decisions after thorough information gathering), hypervigilance (rushed and anxious decision-making), and buckpassing (deferring decisions to others). 
We examine whether these decision-making patterns affect peoples' engagement with AI-generated information in decision-making. In an online experiment with 810 participants tasked with distinguishing food facts from myths, we found that a higher buckpassing tendency was positively correlated with the likelihood of seeking AI information and reported reliance on AI, while being negatively correlated with the time spent reading AI explanations. In contrast, the higher a participant tended towards vigilance, the more carefully they scrutinized the AI's information, as indicated by an increased time spent looking through the AI's explanations. These findings suggest that a person's decision-making pattern plays a significant role in their interactions with AI suggestions, which provides a new understanding of individual differences in AI-assisted decision-making. 
\end{abstract}



\received{20 February 2007}
\received[revised]{12 March 2009}
\received[accepted]{5 June 2009}

\maketitle

\section{Introduction}
Conversational AI tools like ChatGPT provide new ways for individuals to obtain information that can support their decision-making. With their widespread availability and advanced capabilities, people increasingly use these tools to ask for nutrient suggestions~\cite{niszczota2023credibility}, medical advice~\cite{liu2023using}, or personal financial recommendations~\cite{hassani2023role}.
Despite their extensive capabilities, current conversational AI tools supported by large language models (LLMs) have been found to generate inaccurate and biased information in various domains including politics, law, and medicine~\cite{bubeck2023sparks,10.1093/jla/laae003,guerreiro2023hallucinations,tao2023auditing,rozado2023political,niszczota2023credibility}. Reliance on these tools without caution could therefore lead to misinformed decisions and contribute to the spread of misinformation~\cite{feng-etal-2023-pretraining}.

To mitigate these risks and inform interventions, research in Human-AI interaction (HAI) has focused on factors related to AI systems that affect users' reliance, such as accuracy and transparency ~\cite{kim_humans_2023,cao_designing_2024,zhang_effect_2020,wang_are_2021,schemmer_appropriate_2023,schoeffer2024explanations,wang_watch_2023}. 
Other research has explored how individual differences are correlated with varying behaviors in HAI. For example, researchers have shown that attitudes toward, and reliance on, AI vary across different demographics and personality traits~\cite{lai_human_2019,bucinca_trust_2021,feridun2024roles,MA2023102362}. However, a research gap remains in understanding how people's decision-making patterns might influence their use of and reliance on AI.  

To address this gap, this paper explores how individuals' decision-making patterns impact their tendency to rely on seeing knowledge offered by an AI---in our case ChatGPT. As such, we conceptualize individuals' reliance on AI as their engagement with AI-generated information when AI suggestions are available. When decisions have to be made, who seeks out the suggestions offered by ChatGPT? Who is more susceptible to this information despite the risks? And how are these choices related to individuals' decision-making patterns?

Our work builds on a well-established decision-making framework proposed by psychologists Irving Janis and Leon Mann~\cite{mann_melbourne_1997}, which has been widely validated with individuals from various countries~\cite{Cotrena2017AdaptationAV,de2004decision,colakkadioglu2015study,bailly2011adaptation,nota2000adattamento}. Under this framework, decision-making patterns are coping patterns that correspond to situational stress. Adoption of these patterns varies among individuals since people cope with decision-related stress differently, from vigilantly searching for information on their own to deferring decision-making to others. These decision-making patterns can be closely mapped to current interactions with AI systems: the amount of information individuals search during decision-making may impact how much information they need when evaluating AI suggestions; whether individuals defer decision-making to others in \textit{human-human} interactions could influence their reliance on AI during \textit{human-AI interactions}. 

To explore whether individual decision-making patterns predict a person's AI-assisted decision-making tendencies, we preregistered\footnote{ 
 See our preregistration on the Open Science Framework: \url{https://osf.io/z9527/?view_only=ea63631787a74b3c8edd7a8ddc2dbd6b}} and conducted an online experiment (n=810) in which we asked participants to decide whether statements about food are facts or myths, letting them choose whether they wanted to seek ChatGPT's decisions and explanations. To increase the stakes, participants were told to put together a pamphlet that should only include factual nutrition statements, and to imagine that the pamphlet will be shared with an organization that focuses on improving children's health.

Our results show that people differ in their interaction with ChatGPT depending on their individual decision-making pattern. Concretely, the higher a participant's tendency to be a vigilant decision-maker, the more time they spent looking through ChatGPT's explanations, underlining their tendency to scrutinize information before deciding. Conversely, the higher a participant's tendency for buckpassing, the more likely they were to choose to see ChatGPT's decisions, the less time they spent looking at its  explanations for the decision, and the higher their self-reported reliance on ChatGPT. These findings indicate that, in general, buckpassers are less likely to question information provided by an AI, making them more susceptible to misinformation in AI responses than decision-makers who score low on buckpassing or high on vigilance. 

Overall, our work contributes (1) new empirical insights into how people differ in their interactions with AI, implying that parts of the population may be more vulnerable to the risks of AI; (2) design implications for AI tools that go beyond one-size-fits-all, including suggestions for leveraging the benefits of AI technologies while mitigating their risks; (3) a new, publicly available dataset (available at the GitHub repository\footnote{\url{https://github.com/Mooniem/passing_the_buck_to_AI}}) with the demographics, decision-making patterns, and AI interactions of 810 participants, which can be used by researchers for replicating and extending our analyses.  

\section{Related Work}

In this section, we first describe prior studies on AI-assisted decision-making to provide some background on common terminology. This section also sets the stage for our subsequent exploration of factors that have been found to influence human-AI interaction---in particular trust and reliance---and how decision-making patterns may be a missing, but important piece in the puzzle. 

\subsection{Studies of AI-assisted Decision-Making and Terminology}
AI-assisted decision-making, also known as human-AI decision-making, broadly refers to scenarios where AI provides suggestions to help humans make decisions~\cite{10.1145/3610219, lai_towards_2023}. To investigate the effectiveness of providing AI suggestions, recent research has examined users' trust and reliance behaviors such as \textit{overall AI reliance}, \textit{appropriate reliance}, \textit{over-reliance}, and \textit{under-reliance}. Overall AI reliance measures users' reliance behaviors in AI-assisted decision-making regardless of the desirability of such reliance. Users' trust and reliance on systems are often captured via objective measures such as the agreement of an individual's final decision and an AI's suggestion in decision-making tasks~\cite{papenmeier2022s,lu_does_2024,cao_understanding_2022,cao_how_2023,yin2019understanding,morrison_evaluating_2023} or switching from one's decision to the suggestion of an AI~\cite{yin2019understanding,morrison_evaluating_2023,cao_how_2023}. While behavioral metrics were also used to quantify individuals' trust in AI systems~\cite{yin2019understanding}, later researchers pointed out that trust and reliance are distinct concepts with different properties~\cite{lee2004trust}. Trust is considered to guide yet not necessarily determine reliance behaviors. Thus these behavioral metrics are often used to measure reliance, and separate trust-related questionnaires are used to measure trust in AI systems. Subjective measures such as perceived reliance are also often used in prior studies~\cite{cao_understanding_2022,bucinca_trust_2021}. Generally,  \textit{appropriate reliance} is desired because it means that users only adopt the AI's suggestions if they are correct. In contrast, \textit{over-reliance} (when users adopt the AI suggestions despite it being inaccurate \cite{bucinca_trust_2021}) and \textit{under-reliance} (where users fail to adopt the correct AI suggestions~\cite{vasconcelos_explanations_2023}) are undesirable. Notably, studies examining these user behaviors often employ a two-stage decision-making workflow in which AI suggestions are integrated into the collaboration process, by automatically presenting the AI suggestions before or after the users make their decisions \cite{10.1145/3610219,morrison_evaluating_2023,10.1145/3411764.3445522,yin2019understanding,cao_how_2023,lu_does_2024}. As such, the measurement of reliance is often limited to the  experimental design that focuses on agreement with AI's decision or switching. However, to examine whether AI suggestions are effective, it is also critical to understand whether and how people consider them, which might not be universal among individuals. Thus, our study extends the prior behavioral metrics for reliance and takes into account whether and how people choose to see AI suggestions. 
\subsection{Factors that Influence Trust and Reliance on AI}
Recent studies find variance in individuals' trust in, and reliance on AI suggestions and several factors related to system design and individual characteristics. In terms of system-related factors, evidence from empirical studies identified the effects of design and performance of AI systems such as the accuracy of the AI and whether the AI provides a reason for its decision \cite{lai_human_2019,wang_watch_2023,bucinca_trust_2021,schemmer_appropriate_2023,schoeffer2024explanations}. Regarding user characteristics, evidence from survey studies suggests that users' attitudes towards AI systems vary across demographics, including age, education levels, and knowledge about AI~\cite{gillespie_trust_2023,MA2023102362,araujo_ai_2020,chong_human_2022}. 
For example, in a study on individuals' acceptance of ChatGPT, older individuals are found to be less willing to accept ChatGPT compared to younger individuals~\cite{MA2023102362}. \citet{araujo_ai_2020} find that education levels and knowledge about AI and algorithms are positively associated with users' perceived usefulness and fairness of AI. In addition, personal traits have also been found to play a role in users' adoption of suggestions from AI systems: \citet{chong_human_2022}, for instance, find that self-confidence directs individuals' decisions to accept or reject suggestions from the AI. 
Focusing on individuals' Big-Five personality traits (including openness, conscientiousness, agreeableness, neuroticism, and extraversion), \citet{cai2022impacts} identify that individuals who score high in conscientiousness---a tendency related to self-control and responsibility---have higher trust in AI systems that offer both human-requested and system-initiated suggestions. ~\citet{10.1145/3628454.3629552} show that students from India and Thailand who score high in openness (being open to new experiences) are more likely to use ChatGPT. Meanwhile, they find negative effects of agreeableness (being cooperative, friendly) and neuroticism (high tendency toward negative feelings) on students' usage of ChatGPT. Our study extends this line of work by focusing on the effects of individual decision-making patterns in AI-assisted decision-making. While decision-making patterns are found to relate to decision self-esteem and have points of contact with individual differences on personality dimensions such as self-efficacy~\citep{mann_melbourne_1997}, they hold distinct value in human decision-making theory. Unlike personality dimensions that are stable traits, these patterns are individuals' coping strategies with situational stress, which can be influenced by context, offering value for design interventions that help adapt to individuals' decision states.

\subsection{Decision-Making Theories and Individual Patterns}
Scholars from various social science disciplines---such as economics and psychology---have developed different theories explaining the mechanisms behind individuals' decision-making. The rational choice theory, rooted in economics, posits that individuals calculate the utilities of all options to make decisions. Challenging this proposition, \citet{simon_behavioral_1955} argues that people often seek satisfactory rather than optimal choices due to their limited capacity to evaluate numerous options. \citet{schwartz2002maximizing} propose that people tend to either have a tendency of maximizing or satisficing: maximizers tend to assess options carefully and choose one that provides maximum benefit later on, while satisficers tend to settle for an acceptable decision. Different from the economists' perspectives that focus on the utility of decisions, psychologists develop constructs and models of decision-making that relate to motivational processes, cognitive styles, and personality dimensions and traits. The motivational processes refer to decision-making processes in response to certain motivations such as preserving self-image or dealing with negative thoughts and feelings. Example constructs for cognitive styles include need for cognition (the tendency to exert greater cognitive effort to evaluate information)~\cite{cacioppo1982need}. \citet{mann_melbourne_1997}'s theory takes into account individuals' emotions and stress-coping patterns (motivational processes) in their decision-making process. Their four decision-making patterns are derived from a survey study and are described as (i) vigilance (making decisions only after a comprehensive search of information), (ii) hypervigilance (approaching decisions in a hurried and anxious way), (iii) buckpassing (avoiding decisions or deferring to others), and (iv) procrastination (putting off the decision). See  Table~\ref{table:decision_patterns} for an overview and example scale items.

According to \citet{Janis1977DecisionMA},  all four decision-making patterns are in the repertoire of individuals, but people tend to rely on one of these patterns more than on others.  Importantly, \citet{mann_melbourne_1997}'s decision-making patterns have been widely validated with participants from various countries and demographics~\cite{Cotrena2017AdaptationAV,de2004decision,colakkadioglu2015study,bailly2011adaptation,nota2000adattamento} and have been shown to be relevant in real-world decision-making scenarios~\cite{brown2016decision,alexander2017reported,kim_nowhere_2022}. Moreover, \citet{mann_cross-cultural_1998} reveal that individuals' reliance on decision-making patterns (especially buckpassing) varies between East Asian and Western cultures. Together these studies suggest that people vary in their approach to decision-making. They also found that individuals' decision self-esteem is positively correlated with vigilance and negatively with hypervigilance and buckpassing.

Decision-making patterns have been found to affect individuals' decision outcomes. 
In the context of economic choices, research shows that when given a time constraint,  maximizers tend to browse more options and change their decisions before making the final purchase than satisficers~\cite{chowdhury_time-harried_2009}. 
Adopting \citet{mann_melbourne_1997}'s decision-making categories, \citet{kim_nowhere_2022} found that individuals' decision-making patterns are linked to their quality of life, including physical and psychological. These findings shed light on the impact of decision-making patterns on individuals' lives, which could also be manifested in the context of AI-assisted decision-making. \citet{jugovac2018investigating} has examined how maximizers and satisficers differ in their interactions while presenting AI recommendations; however, they found no effect. Notably, their study design focuses on participants' interactions with two provided recommendations instead of a design setting where relying on an AI is optional. Our study further explores this line of work in a design setting where AI support is optional, examining how users' behaviors differ when they are faced with uncertain information and have the option to defer to the AI.  
\section{Methodology}
\subsection{Hypotheses}
We developed our hypotheses based on the decision-making patterns by \citet{mann_cross-cultural_1998} (summarized in Table~\ref{table:decision_patterns}). Importantly, these decision-making patterns emphasize whether people delegate decision-making to others and how they search for information under decision stress. As such, we believe these decision-making patterns are likely predictive of behaviors that we commonly see when humans interact with an AI, such as relying on an AI to seek further information or overrelying on an AI to avoid making one's own decision.

\begin{table}[h!]
\centering
\caption{Decision-Making Patterns, Descriptions, and Example Statements from the Melbourne Decision-Making Questionnaire}
\begin{tabular}{l|m{6cm}|m{4cm}}
\toprule
\hline
\textbf{Decision-Making Pattern} & \textbf{Description} & \textbf{Example Item} \\ \hline
Vigilance & This pattern refers to a rigorous information search, such as when a person tries to evaluate different alternatives before making a decision. & ``I consider how best to carry out the decision.'' \\ \hline
Hypervigilance & This pattern refers to a frantic way of making decisions to relieve emotional stress. & ``After a decision is made I spend a lot of time convincing myself it was correct.'' \\ \hline
Buckpassing & This pattern refers to a tendency to leave decisions to someone else. & ``I prefer that people who are better informed decide for me.'' \\ \hline
Procrastination & This pattern refers to a tendency to put off making decisions. & ``I waste a lot of time on trivial matters before getting to the final decision.'' \\ \hline
\bottomrule
\end{tabular}
\label{table:decision_patterns}
\Description{
The table summarizes four decision-making patterns, including their descriptions and example items:  

1. Vigilance: Describes a careful information search to evaluate alternatives before deciding. Example: "I consider how best to carry out the decision."  
2. Hypervigilance: Refers to a frantic decision-making style driven by emotional stress. Example: "After a decision is made I spend a lot of time convincing myself it was correct."  
3. Buckpassing: Involves deferring decisions to others. Example: "I prefer that people who are better informed decide for me."  
4. Procrastination: Involves delaying decisions and focusing on trivial matters. Example: "I waste a lot of time on trivial matters before getting to the final decision." }
\end{table}

Overall, we hypothesize that decision-making patterns could affect individuals' interactions with AI suggestions. Specifically, by \textbf{AI suggestions}, we refer to the recommended decisions from the AI (\textbf{AI decisions}) and its text-based justifications for its recommendation (\textbf{AI explanations}). Our study imagines scenarios where people encounter information online that may be right or wrong and use an AI tool (e.g., ChatGPT) to assess its credibility. When using an AI, people can seek the AI's decision and, if needed, the AI's explanations for the decision. 
We speculate these AI suggestions are used differently across individuals with varying decision-making patterns, as outlined in Table~\ref{tab:decision_hypotheses}. For vigilant decision-makers who tend to conduct a comprehensive search of information and evaluate it rigorously, we hypothesize that they are more likely to seek the AI's suggestions for additional information even if they already have an answer. However, as they also tend to be more confident in their own decisions~\cite{mann_cross-cultural_1998}, they might not necessarily prioritize the AI's decision over their own. 

People who exhibit hypervigilant decision-making behaviors tend to search for information frantically, but not rigorously, before arriving at a decision. Thus, we hypothesize they will quickly seek information from the AI (i.e., see AI decision and explanations) without spending much time evaluating the information. As they experience high emotional stress and are less confident in their decisions~\cite{mann_cross-cultural_1998}, they would be more likely to rely on the AI's suggestions to reach their decisions.

We further hypothesize that people who exhibit a buckpassing tendency will seek out the AI's suggestions to avoid making any decisions themselves, leading to a high reliance on AI. As a result, they would not consider AI's suggestions as much, leading to less time spent on AI explanations.

\begin{table}[h!]
\renewcommand{\arraystretch}{1.4}
\caption{We develop our hypotheses based on individuals' decision-making patterns, focusing on their behavioral characteristics related to information search.}
\centering
\begin{tabular}{lp{7.2cm}|l}
\hline
\textbf{Decision Making Pattern} & \textbf{Hypotheses} & \textbf{Notation} \\ \hline
\multirow{3}{*}{The higher people score on vigilance,} 
    & the \textit{more} likely they are to choose to \seeAIdecision.  & H1a  \\ 
    & the \textit{more} time they spend \seeAIexplanations. & H1b   \\ 
    & the \textit{lower} they report their \perceivedAIreliance to be.  & H1c  \\ \hline
\multirow{3}{*}{The higher people score on hypervigilance,} 
    & the \textit{more} likely they are to choose to \seeAIdecision. & H2a  \\ 
    & the \textit{less} time they spend \seeAIexplanations.  & H2b \\ 
    & the \textit{higher} they report their \perceivedAIreliance to be. & H2c \\ \hline
\multirow{3}{*}{The higher people score on buckpassing,} 
    & the \textit{more} likely they are to choose to \seeAIdecision.  & H3a \\ 
    & the \textit{less} time they spend \seeAIexplanations.  & H3b \\ 
    & the \textit{higher} they report their \perceivedAIreliance to be. & H3c \\ \hline
\end{tabular}
\label{tab:decision_hypotheses}
\end{table}


We did not include \textit{procrastination} in our hypotheses because this would require leaving participants with potentially infinite time, which we did not deem feasible. However, procrastination has been shown to be correlated with behaviors or buckpassing and hypervigilance~\cite{bouckenooghe2007cognitive}, and all three decision-making patterns are considered forms of defensive avoidance \cite{Janis1977DecisionMA}. 

\subsection{Experiment Design}\label{sec:experiment-design}
To examine our hypotheses, we designed an online study (preregistered at 
 \hyperlink{https://osf.io/z9527/?view_only=ea63631787a74b3c8edd7a8ddc2dbd6b}{link}) that includes (i) a questionnaire and (ii) a task requiring participants to decide on the factuality of a series of nutrition statements while having the option to seek out AI suggestions (\textbf{AI decisions} and \textbf{AI explanations}). We generated AI decisions and explanations using ChatGPT before the study (rather than having participants interact with ChatGPT live) to ensure consistency of response quality. The study was approved by our Institutional Review Board and launched in English on the volunteer-based online study platform LabintheWild\footnote{\url{https://labinthewild.org/}}. We encouraged honest participation by informing participants in advance that they would receive personalized result---their own decision-making patterns and performance on evaluation food facts and myths---at the end of the study. 
\vspace{2mm}
 
\noindent \textbf{Nutrition Statements for the Main Task:} We used 30 nutrition statements and their accuracy (i.e., fact or myth) from~\citet{florenca_food_2021} who curated a list of popular food (mis)-conceptions from online sources. In their study with 503 participants, several of these statements were inaccurately classified as facts or myths by participants, and even by participants working in areas related to nutrition, suggesting that the statements have a range of difficulty. 
Among the 30 statements, 9 nutrition statements are facts and 21 are myths. One example false statement (a myth) is ``Pregnant women should be eating for two'', while an example true statement (a fact) is ``Dairy products should be consumed in between two and three portions
per day''. The full list of statements with fact/myth labels is in the Supplementary Materials. \\ [-2mm]

\noindent \textbf{AI Suggestions (AI Decisions and Explanations):} We generated AI decisions and explanations for each nutrition statement using ChatGPT\footnote{\href{https://chat.openai.com/}{https://chat.openai.com/}} in July 2023, prompting ChatGPT in the format of ``\texttt{[Nutrition Statement]}.\texttt{ Is this fact or myth?}'' After the model generated an answer to the prompt, we asked it to shorten its answer to less than four sentences by prompting ``\texttt{Rewrite it in less than 4 sentences.}'' Based on nutrition statements from~\citet{florenca_food_2021},  ChatGPT produced incorrect answers for three out of the 30 statements (two false facts and one false myths). We used ChatGPT's decision whether the statement is a fact or a myth as the \textit{AI Decision} and the rest of the information it provided as the \textit{AI Explanation}.

We then chose 21 accurate AI suggestions (each comprising a decision and explanation), the 3 inaccurate AI suggestions that ChatGPT provided, and further inverted 6 suggestions to become false, resulting in 21 accurate and 9 inaccurate AI suggestions. From these two pools of statements, we randomly sampled 6 statements with accurate AI suggestions and 4 with inaccurate AI suggestions. We included more accurate than inaccurate AI decisions in our experiment to be faithful to the real-world performance of AI models (i.e., ChatGPT provided accurate responses to most of the 30 nutrition statements). To generate inaccurate AI explanations for these 6 statements, we prompted ChatGPT with \texttt{``What are some arguments people would use to support this statement as a fact/myth: [Nutrition Statement]}.''~and also asked it to shorten its answer to have fewer than four sentences.

\subsection{Study Procedure}
Our online study, directed at intrinsically motivated volunteer participants, was advertised with the question ``What is your decision-making style?" Participants were able to receive personalized results at the end of the study.  As they entered the study page, we explained the goal of the experiment and that they would be asked to judge a set of nutrition statements. Participants could proceed to the experiment only if they consented. Participants were then asked to fill out a demographic questionnaire including questions about their age, gender, education levels, and country. We also inquired about their knowledge of nutrition (``How would you describe your knowledge of nutrition compared to the general public?'' with the answer options being: ``Very limited'', ``Limited'', ``Average'', ``Above average'', and ``Expert'') and assessed their perception of the accuracy of AI-generated information (``To what extent do you believe that conversational AI (e.g. ChatGPT) provides correct information?'' with answer options ``I don't know what ChatGPT is'', ``It never provides anything correct'', ``It provides correct information sometimes'', ``It provides correct information usually'', and ``It always provides correct information.''). 

Next, participants were asked to complete the Melbourne Decision-Making Questionnaire (MDMQ)~\cite{mann_melbourne_1997} (see Supplementary Materials), with 10 questions being shown before the main tasks and the remaining 7 questions being shown after the main task (all in random order) to reduce the perceived questionnaire length. 

\begin{figure*}[t!]
    \centering
    \includegraphics[width=\textwidth]{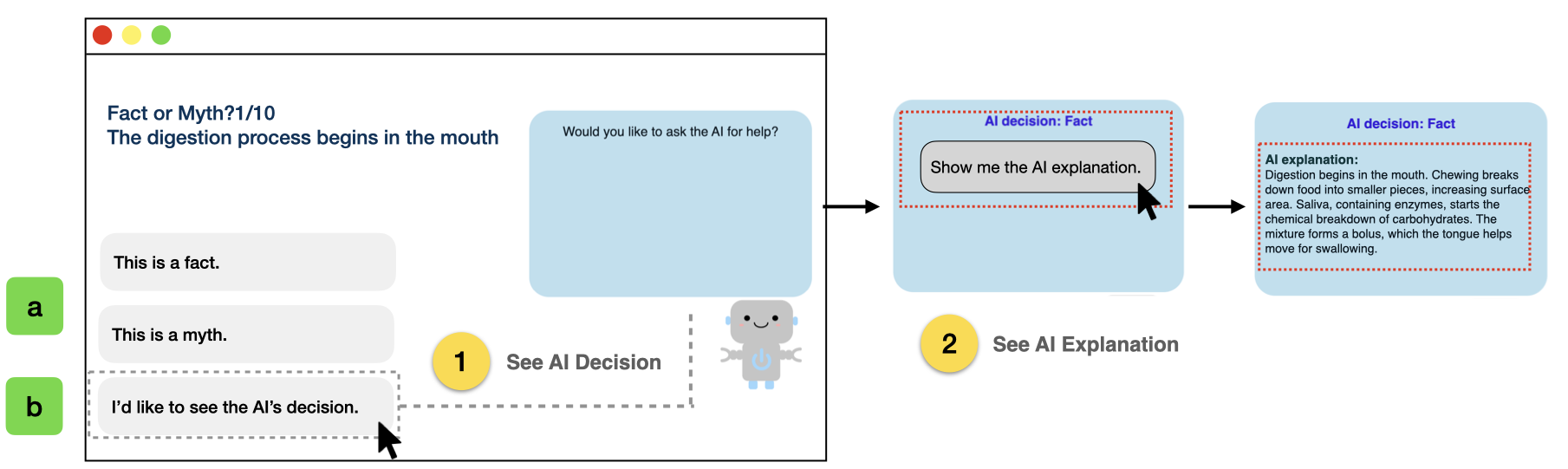}
    \caption{The study interface and workflow for each statement. Participants first read one statement at a time. They could then decide whether the statement is a fact or myth (\plabel{a}). If participants decided at this point, they could move on to the next statement. Alternatively, participants could choose to reveal the AI's decision (\ilabel{1}) showing ChatGPT's decision (\plabel{b}). Participants could then decide to choose fact or myth or use the AI decision. They could also choose to see AI's explanation to reveal the AI explanation before making their decision (\ilabel{2}). }
    \label{fig:workflow}
    \Description{This figure illustrates the interaction flow of the main task for participants. The figure indicates two possible interaction flows for participants when they are evaluating food facts or myths. One of the interaction flows is that participant can choose to provide their answer based on their own knowlege and select the correponding answer. The other workflow is particpants can choose to select See AI decision button to see AI's decision for the evaluation of statement. After the AI's decision is revealed, participants would be shown a button that says "Show me the AI's explanation". At this point, they can either skip the explanation and just make their final evaluation, or click the button to reveal AI's text-based explanations.}
\end{figure*}

~\\ [-2mm]
\noindent \textbf{Main task:} The main task showed participants 10 nutritional statements (6 accurate and 4 inaccurate decisions, drawn from the larger set of 30 statements using stratified random sampling). For each statement, participants were asked to assess whether it was a fact or myth while being provided options to use AI assistance (see Figure~\ref{fig:workflow}). To encourage participants to make cautious decisions about these statements, we introduced the following scenario:  
\begin{quote}
\textit{Imagine your manager asks you to put together a nutrition pamphlet to share with an organization that focuses on improving children's health. To do so, you will need to evaluate 10 nutrition-related statements and determine if they are facts or myths. You will have access to an AI that can assist you with these assessments. However, please keep in mind that the AI's suggestions may not always be entirely accurate or definitive. Whether to use the AI's decision is completely up to you.}
\end{quote}
We mentioned the limited performance of AI assistance to ensure participants knew that not all information presented to them was factually correct, in line with current state-of-the-art conversational AI systems which do not always output accurate information \cite{deiana2023artificial}. Informing participants about the imperfection of decision aid is often used in prior work assessing users' trust and reliance on automated support systems \cite{dzindolet2003role}.  While it might prime participants to refrain from seeking AI suggestions, it could also encourage participants to critically evaluate AI suggestions rather than accept them blindly.

If participants chose ``I'd like to see the AI decision,'' they were provided two additional options: 
to see an AI explanation or to immediately accept the AI decision (without seeing an explanation). This design was motivated by our hypotheses:  \textit{\seeAIdecision} is designed to observe whether users prefer to refer to the AI's decision before making their own. Subsequently, we offered the option to \textit{\toseeAIexplanations} to investigate whether some users might want to further investigate and seek additional details.  

After an AI decision was revealed, participants could select \textit{Use the AI decision} to indicate they accept the AI decision; however, this design revealed its limitations as participants barely selected this option based on our analysis of results which we discuss later. 
For each statement, we recorded the time they took to evaluate the AI's decision and explanations before final decision-making.   

 At the end of the study, participants were asked to 
 assess their \perceivedAIreliance during the study via the question ``How much did you rely on the AI in your decisions for food facts and myths?''. Participants could select from a five-point scale with 1 being `Didn't rely on it at all' and 5 being `I relied on it all the time'. 

The experiment ended with a page showing participants their decision-making patterns, how many nutrition statements they judged correctly versus incorrectly, and which statements they got wrong.  
We further debriefed participants by stating that we used ChatGPT for the AI-generated content during the experiment and that only some of the nutrition statements and explanations were correct. 



\subsection{Metrics}
We operationalize individuals' interactions with AI suggestions via a combination of clicking behaviors, reading time, and self-reported reliance:
\begin{itemize}
     \item \textbf{See AI Decision} is a binary variable that indicates whether participants select the option ``I'd like to see the AI's decision'' (\ilabel{1} in~\autoref{fig:workflow}) for each statement. This variable measures whether participants seek more information from the AI during the task. 
     For each participant, we also calculated the percentage of statements they see AI decisions (\textbf{Frequency of See AI Decision}).
    \item \textbf{See AI Explanations} is a binary variable indicating whether participants selected the option ``Show me the AI explanation'' (\ilabel{2} in~\autoref{fig:workflow}) for each statement. This measures whether people seek more information from the AI in the task after seeing the AI decision. For each participant, we also calculated the percentage of statements they see explanations out of the 10 statements they see (\textbf{Frequency of See AI Explanations}).
    \item \textbf{Time Spent Seeing AI Explanations} is a continuous variable that measures the time it takes participants to select a final evaluation of a statement after they choose to see the AI explanation. We only examined this variable for participants who select a response after they click ``See AI Explanation''. 
    \item \textbf{Perceived AI Reliance} measures individuals' self-reported reliance on AI assistance in the study based on their responses to the question ``How much did you rely on the AI in your decisions for food facts and myths?'' We coded levels 1 to 5 (1 for ``Didn't rely on it at all'', 5 for ``I relied on it all the time''). We adopted a one-item measurement approach to examine individuals' perception-based reliance on AI based on prior study design~\cite{cao_understanding_2022}. The item was adapted to focus on our nutrition task, assessing participants' reliance explicitly during the study.
\end{itemize}
 We examine the effects of decision-making patterns while considering the demographic covariates (age, gender, education levels) as well as the following variables: 
\begin{itemize}
    \item \textbf{Vigilance, Hypervigilance, Buckpassing}: Based on participants' answers to the \textbf{Melbourne Decision Making Questionnaire (MDMQ)}~\cite{mann_melbourne_1997} on a 3-point scale (``True for me'' (score 2), ``Sometimes true for me'' (score 1), ``Not true for me'' (score 0)), we obtained the total score for each of their decision-making dimensions. Each decision-making pattern has its own subscale, consisting of a unique set of statements distinct from those used in the other subscales. Therefore, while we excluded procrastination in our hypotheses and did not administer that portion of the MDMQ, this should not impact the validity of measuring buckpassing, vigilance, and hypervigilance. The maximum score is 10 for hypervigilance, and 12 for both vigilance and buckpassing. The vigilance and buckpassing subscales consist of 6 items each, while the hypervigilance subscale
consists of 5 items.
    \item \textbf{Domain Knowledge} is based on participants' self-reported knowledge level of nutrition compared to the average public. We categorized participants into three groups (``Below Average'', ``Average'', ``Above Average'')  due to the imbalance of individuals among the five subgroups (``Very Limited'' (41), ``Limited'' (92), ``Average''(351), ``Above Average''(300), ``Expert'' (27)). When participants reported their knowledge levels as ``Above Average'' or ``Expert'', their domain knowledge was coded as ``Above Average''. We categorized ``Very Limited'' and ``Limited'' to be ``Below Average''.  Note that we treated this variable as nominal since we were uncertain about whether the increase of likelihood to see AI suggestions and perceived AI reliance would be the same from ``Below Average'' to ``Average'' compared to from ``Average'' to ``Above Average''.
    \item \textbf{Perception of AI} of participants were coded as ``High'' and ``Not high'' based on their perception of the accuracy of conversational AI before participating in the experiment. If they selected ``It provides correct information usually'' or ``It always provides correct information'', it was coded as ``High'', otherwise it was coded as "Not high". 
\end{itemize}

\subsection{Participants}
Participants in our study were 917 volunteers recruited via the online study platform LabintheWild between August 2023 and April 2024. The platform was chosen to allow us to recruit diverse participants of various ages and education levels to ensure a broad spread of decision-making styles. Participation on LabintheWild is open to anyone without signing up; we therefore obtained a waiver of parental consent from our IRB so that minors were able to take our study. To reduce the risk of participants mistaking factually incorrect nutrition information as correct, the study instructions clearly stated that not all nutrition statements are factual and that the goal is to test how much participants know about nutrition. As part of this debrief,  participants were also told which of the nutrition statements they wrongfully assumed to be true at the end of the experiment. 

We excluded any participants who did not complete the study or affirmed they had taken the study before. To reduce the risk of including participants who were not truthfully responding to the task (i.e., those exhibiting satisficing behavior~\cite{kapelner2010preventing}), we also excluded participants who consistently selected either the option ``True for me'' or ``Not true for me'' for all questions in the decision-making questionnaire. After examining the distribution of participants' total time spent completing the study (included in the Appendix), we decided to remove participants whose study completion times were outliers---either extremely short or extremely long---and unlikely to represent a serious attempt at completing the tasks. On average, participants took 9 minutes to finish the study (Median = 468 seconds, SD= 297 seconds).  Thus, we eliminated participants who completed the study in less than 240 seconds minutes (N=18) or more than 1200 seconds (N=54), accounting for participants who did not seek AI suggestions and those who reviewed all suggestions. In total, we excluded 107 participants, resulting in a final sample of size 810. 

The final set of participants reported being from 69 countries, with the majority from the United States (44\%), followed by the United Kingdom (9\%), Canada (8\%), India (5\%), Germany (3\%), and Australia (3\%, all others $\leq 3\%$). 
Half of our participants reported  pursuing or having obtained a college education, approximately 29\% of participants reported pursuing or obtaining  a high school education; with the rest reporting to pursue or having  obtained a graduate education.
Our study participants were between 12 to 90 years of age ($M =31$, $SD =14$).   
A detailed breakdown of demographics for participants is shown in Table ~\ref{tab:dm_stat}. 

\begin{table*}[t!]
\small
\centering
\caption{Sample sizes of study participants across different demographics in terms of participants' education levels, gender, age group, self-reported knowledge of nutrition, and their perception of the accuracy of AI-generated information.}
\begin{tabular}{lr}
\hline
 & N \\
\multicolumn{2}{l}{\textbf{Overall}}\\
\hline
\hspace{1em}Overall & 810 \\

\multicolumn{2}{l}{\textbf{Education}}\\

\hspace{1em}<=High School Education & 231 \\

\hspace{1em}Pursuing or Have obtained college education & 403 \\

\hspace{1em}Pursuing or Have obtained graduate education & 176 \\
\hline
\multicolumn{2}{l}{\textbf{Gender}}\\

\hspace{1em}Female & 409 \\

\hspace{1em}Male & 363 \\

\hspace{1em}Non-binary/No-disclosure & 38 \\
\hline
\multicolumn{2}{l}{\textbf{Age Group}}\\

\hspace{1em}Less than 18 & 106 \\

\hspace{1em}18-24 & 264 \\

\hspace{1em}25-34 & 189 \\

\hspace{1em}35-44 & 107 \\

\hspace{1em}45-55 & 82 \\

\hspace{1em}Above 55 & 62 \\
\hline
\multicolumn{2}{l}{\textbf{Nutrition Knowledge}}\\

\hspace{1em}Very limited & 41 \\

\hspace{1em}Limited & 92 \\

\hspace{1em}Average & 351 \\

\hspace{1em}Above Average & 300 \\

\hspace{1em}Expert & 26 \\
\hline
\multicolumn{2}{l}{\textbf{Perception of AI}}\\

\hspace{1em}I don't know what conversational AI (ChatGPT) is. & 64 \\
\hspace{1em}It never provides anything correct. & 8 \\
\hspace{1em}It provides correct information sometimes. & 346 \\
\hspace{1em}It provides correct information usually. & 378 \\
\hspace{1em}It always provides correct information. & 14 \\
\hline
\end{tabular}
\Description{This table shows the sample sizes of study participants across different demographics, including their education levels, gender, age group, self-reported knowledge of nutrition, and their perception of the accuracy of AI-generated information. The table contains columns for the different categories and the corresponding sample sizes for each.}
\label{tab:dm_stat}
\end{table*}
~\\ [-6mm]

\section{Analyses and Results}\label{analysis-section}
\subsection{Descriptive Statistics}
This section presents our sample characteristics and key variables, as a form of robustness check in line with prior work \citet{mann_cross-cultural_1998} and \citet{florenca_food_2021}. Then we present detailed analysis results on (1) Effect of Decision-Making Patterns on Seeing AI Decisions (H1a, H2a, H3a), Effect of Decision-Making Patterns on Seeing AI Explanations (H1b, H2b, H3b), and Effects of Decision-making Patterns on Perceived AI Reliance (H1c, H2c, H3c).

\paragraph{Variance in Decision-Making Patterns and Measurement Reliability}
The full summary statistics of decision-making patterns for the different demographics are shown in the Appendix (Table~\ref{dm_stats_table}). On average, our study participants scored 9.41 ($SD =2.31$) out of 12 on vigilance, 4.25 ($SD =2.61$) out of 10 on hypervigilance, and 4.56 ($SD=3.03$) out of 12 on buckpassing. 
Each subscale reaches a satisfactory level of internal reliability. Cronbach's alphas for each subscale is 0.74 (vigilance), 0.81 (buckpassing), and 0.74 (hypervigilance).
\footnote{Cronbach's alpha measures internal consistency and reliability of a scale, ranging from 0 to 1. Values from 0.70-0.90 are recommended to be acceptable values \cite{tavakol_making_2011}.} Cronbach's alphas were 0.80,0.87,0.74 from \citet{mann_cross-cultural_1998}. The mean and standard deviation of our measures fell within a similar range compared to those observed in \citet{mann_cross-cultural_1998} across different countries. Participants' hypervigilance had significantly positive correlation with their measure of buckpassing ($r = 0.59, p<0.001$), which aligns with prior studies \cite{filipe_validation_2020,mann_melbourne_1997}. Additional factor analysis confirmed that these three different decision-making patterns were identified with the same corresponding question items from \citet{mann_cross-cultural_1998}. Details regarding factor loadings and a corresponding diagram are included in the Appendix. These similarities suggest the reliability of our measure of decision-making patterns and the robustness of our samples. 


\paragraph{Variance in Participants' Frequency of Seeing AI Decisions and Explanations}
We first verified if there was variance in participants' interactions with AI suggestions before we identified the respective underlying factors. We found that the frequency of seeing AI suggestions varied across participants, with the average participant choosing to see the AI's decision $30\%$ of the time ($ M= 30\%, SD =26\%, Min= 0\%, Max = 100\%$) and the AI's explanations $23\%$ of the time ($M= 23\%, SD = 23\%, Min= 0\%, Max = 100\%$). 


\paragraph{Participants' Overall Performance} 
On average, participants accurately evaluated 70\% of the statements ($M = 70\%, SD=14\%$).  Participants who scored high in buckpassing (M = $68\%$) had a slightly lower accuracy than those who scored average (M = $70\%$) and low in buckpassing (M = $71\%$). When AI decisions were incorrect, participants scored much lower in accuracy ($M = 23\%$) when they chose to see AI decisions compared to times when they did not ($M =57\%$). When AI decisions were correct, participants scored slightly higher in accuracy when they chose to see AI decisions ($M = 90\%$) compared to times when they didn't ($M = 84\%$). A detailed breakdown of accuracy across different conditions and buckpassing level is included in the appendix  (Figure~\ref{fig:accuracy_breakdown}).  Performance did not significantly differ across the different decision-making styles. 
The performance of our participants on each statement was significantly correlated with participants in \citet{florenca_food_2021} ($r= 0.58, p<0.001$), indicating that most of our participants exhibit a similar knowledge pattern regarding common food myths and facts. 

\paragraph{Agreement with AI suggestions} 
To examine participants' decision behaviors after they see AI decisions, we measured their agreement rate. Proportion of agreement with AI suggestions is often used as a behavioral metric of reliance in prior studies where AI suggestions are automatically shown~\cite{lu_does_2024,cao_understanding_2022,cao_how_2023}. However, since our study design provided AI suggestions on demand, we measured participants' conditional agreement rate conditioned on whether they chose to see AI decisions. Participants agreed with AI decisions on about 83\% of statements when they chose to see them, compared to 67\% when they did not. Among all participants, around 49\% (395 out of 810) of participants agreed with AI every time they chose to see AI decisions. We also visualized the distribution of participants' individual conditional agreement rates in the Appendix (Figure \ref{fig:agreement_rate}).

\begin{figure}[!h]
    \centering
    \includegraphics[width=0.6\linewidth]{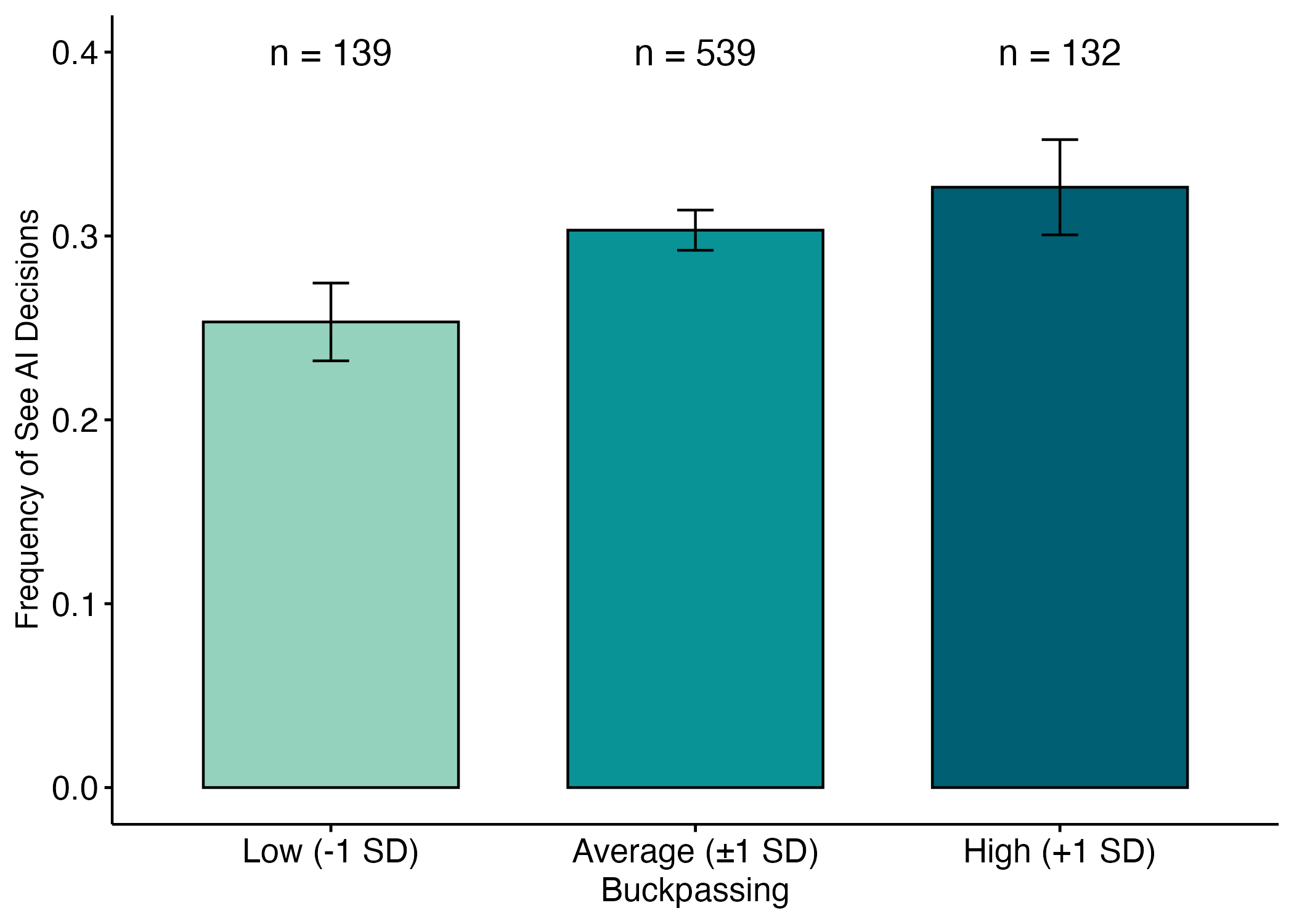}
        \caption{Overview of Frequency of See AI Decisions with Different Levels of Buckpassing: Participants with varying levels of buckpassing show differences in how often they view AI decisions. ``Low'' and ``High'' buckpassing refer to scores that are more than one standard deviation below and above the average, respectively. ``Average'' buckpassing refers to scores that are within one standard deviation of the average. Each bar represents the average frequency, with error bars indicating the confidence intervals. On average, those who score low in buckpassing view AI decisions 25\% of the time, while those who score high view them 33\% of the time.}
    \label{fig:buckpassing_see_AI_decisions}
    \Description{This figure is a bar chart displaying the relationship between buckpassing (a decision-making style) and the frequency of users choosing to "See AI Decisions" across three buckpassing groups: Low, Average, and High. On the y-axis is the average frequency of selecting "See AI Decisions" and the error bar. The scale ranges from 0.0 to 0.4. On the x-axis is the buckpassing levels divided into three categories: low, average, and high. Low level includes participants whose buckpassing score is 1 standard deviation below the mean (marked in red). Average level includes participants with an average buckpassing score. High level includes participants whose buckpassing score is 1 standard deviation above the mean. Sample size for each group are displayed above the bars: Low buckpassing group has n = 139. Average buckpassing group has n = 540.
High buckpassing group has n = 132. The heights of the bars show that participants in the Low and High buckpassing groups selected "See AI Decisions" less frequently than those in the Average group. The error bars are 95\% confidence intervals for the mean frequencies. The Low buckpassing group has a slightly lower frequency with a narrow confidence interval, while the Average and High groups have higher frequencies with overlapping confidence intervals.}
\end{figure}

\subsection{Effect of Decision-Making Patterns on \seeingAIdecision}

To examine the effect of decision-making patterns on individuals' likelihood to see AI decisions (\textbf{H1a, H2a, H3a}), we fit a series of mixed-effects logistic regression models for our binary dependent variables (i.e., whether a participant decided to see the AI decision and the AI explanation for each statement). We included participants' individual decision-making scores (for \textit{vigilance, hypervigilance, buckpassing}) as fixed effects, with participant and nutrition statement as random effects (\textbf{Model 1} in \autoref{tab:regression_results} ) to account for variation across nutrition statements and individual differences. Neither hypervigilance nor vigilance showed a significant effect, suggesting that \textbf{H1a} and \textbf{H2a} are not supported. Buckpassing had a significant positive effect on the likelihood of people seeing AI decisions ($\beta = 0.062, p<0.05$), supporting \textbf{H3a}. We also show an overall frequency of seeing AI decisions across different levels of buckpassing among our participants in \autoref{fig:buckpassing_see_AI_decisions}.

\begin{table}
\caption{Mixed-effects logistic regression results predicting participants' tendency to seek AI suggestions (See AI Decision and See AI Explanations). Regression analysis with only variables related to decision-making patterns indicates a statistically significant effect of buckpassing on one's likelihood to see AI's decision. When including demographic variables, the regression analysis shows that participants' education levels,  self-reported domain knowledge, and perception of AI-generated information have significant effects on participants' tendency to seek AI suggestions.  Reference groups are less than college education, domain knowledge (Above average), and perception of AI (Not high).}
\label{tab:regression_results}
\begin{tabular}{@{\extracolsep{5pt}}lccc} 
\hline 
\hline \\[-1.8ex] 
 & \multicolumn{3}{c}{\textit{Dependent variable:}} \\ 
\cline{2-4} 
\\[-1.8ex] & \multicolumn{2}{c}{See AI Decision} & See AI Explanations \\ 
\\[-1.8ex] & Model 1 & Model 2 & Model 3\\ 
\\[-1.8ex] & Coef. (S.E.) & Coef. (S.E.)& Coef. (S.E.)\\ 
\hline \\[-1.8ex] 
 Buckpassing & 0.062$^{**}$ (0.025) & 0.048$^{*}$ (0.025) & 0.032 (0.026) \\ 
  Hypervigilance & $-$0.001 (0.029) & $-$0.021 (0.029) & $-$0.029 (0.030) \\ 
  Vigilance & 0.015 (0.027) & 0.014 (0.026) & $-$0.007 (0.027) \\ 
  Age &  & $-$0.015$^{***}$ (0.005) & $-$0.016$^{***}$ (0.005) \\ 
  Pursuing or Have obtained college education &  & 0.071 (0.149) & 0.104 (0.153) \\ 
Pursuing or Have obtained graduate education &  & 0.624$^{***}$ (0.185) & 0.631$^{***}$ (0.190) \\ 
  Domain\_Knowledge (Average) &  & 0.594$^{***}$ (0.136) & 0.406$^{***}$ (0.139) \\ 
  Domain\_Knowledge (Below Average) &  & 0.500$^{***}$ (0.187) & 0.108 (0.193) \\ 
  Perception of AI (High) &  & 0.256$^{**}$ (0.121) & 0.080 (0.124) \\ 
  Constant & $-$1.697$^{***}$ (0.315) & $-$1.695$^{***}$ (0.387) & $-$1.704$^{***}$ (0.399) \\ 
 \hline \\[-1.8ex] 
Observations & 8,100 & 8,100 & 8,100 \\ 
Akaike Inf. Crit. & 8,411.871 & 8,379.897 & 7,455.465 \\ 
Marginal $R^2$ / Conditional $R^2$ &	0.006 / 0.470& 0.033 / 0.469 & 0.020 / 0.469 \\
\hline 
\hline \\[-1.8ex] 
\textit{Note:}  & \multicolumn{3}{r}{$^{*}$p$<$0.1; $^{**}$p$<$0.05; $^{***}$p$<$0.01} \\ 
\end{tabular} 
\Description{This image shows the results of a mixed-effects logistic regression analysis predicting participants' tendency to seek AI suggestions. The analysis included three models, each with a different dependent variable: See AI Decision, See AI Explanations, and a combination of the two.

The table displays the coefficient (standard error) for each independent variable in the three models. The independent variables include Buckpassing, Hypervigilance, Vigilance, Age, Pursuing or Have obtained college education, Pursuing or Have obtained graduate education, Domain_Knowledge (Average), Domain_Knowledge (Below Average), Perception of AI (High), and Constant.

The table also provides the number of observations, Akaike Information Criterion, Marginal R^2 / Conditional R^2 values for each model.

The results indicate that participants' education levels, self-reported domain knowledge, and perception of AI-generated information have significant effects on their tendency to seek AI suggestions. Reference groups with less than college education, domain knowledge below average, and perception of AI not high are less likely to seek AI suggestions.}
\end{table}

Additionally, we built a model (\textbf{Model 2}) that controls for demographic covariates since prior research suggests that individuals' demographics factors such as age, education levels, domain knowedge~\cite{kim_humans_2023,MA2023102362,araujo_ai_2020} influence their trust and reliance on AI. 
The details of this model are summarized in \autoref{tab:regression_results}. 
The positive effect of buckpassing remains marginally significant after controlling for participants' age, education, domain knowledge, and perception of an AI ($\beta = 0.048, p=0.06$).
\subsection{Effect of Decision-Making Patterns on \seeingAIexplanations}
To test \textbf{H1b}, \textbf{H2b}, and \textbf{H3b}, we first examined whether participants chose to see AI explanations, which was true for 548 out of 811 participants. Per \textbf{Model 3} in \autoref{tab:regression_results}, there is no significant effect of decision-making patterns on individuals' tendency to see AI explanations. We included the results of the model with only decision-making measures as fixed effects in the Supplemental Materials.

Next, we examined the effect of decision-making patterns on how much time people spend reviewing AI explanations, if they choose to see them. 
On average, participants spent 14 seconds seeing AI explanations ($M= 14.34,SD = 10.73$). We used a linear mixed-effects regression model, considering that each participant could have their own reading speed and each statement had different lengths of AI explanations. Per \autoref{tab:regression_eval_time}, buckpassing had a statistically significant negative effect on individuals' time spent evaluating AI explanations ($\beta= -0.40, p <0.05$). For one additional score of increase in buckpassing dimension, the time users spend on seeing AI explanations decreased by 0.4 seconds. This suggests individuals who score higher in buckpassing tend to spend less time evaluating AI explanations before they make their final decisions. On the contrary, vigilance had a statistically significant positive effect on participants' time spent evaluating AI explanations ($\beta = 0.53,p<0.05$). For one score of increase in vigilance dimension, the time users spent on reading explanations increased by 0.53 seconds. These results support our hypotheses \textbf{H1b} and \textbf{H3b}, but not \textbf{H2b}.

\begin{table}[ht]
\centering
\caption{Linear mixed model predicting individuals' time spent evaluating AI explanations: participants' buckpassing and vigilance tendency significantly affect the time they spent reading AI's explanations if they chose to see them.  
} 
\label{tab:regression_eval_time}
\begin{tabular}{@{\extracolsep{5pt}}lc} 
\\[-1.8ex]\hline 
\hline \\[-1.8ex] 
 & \multicolumn{1}{c}{\textit{Dependent variable:}} \\ 
\cline{2-2} 
\\[-1.8ex] & Time Spent Seeing AI Explanation (Seconds) \\ 
\hline \\[-1.8ex] 
 Buckpassing & $-$0.401$^{**}$ (0.197) \\ 
 Vigilance & 0.528$^{**}$ (0.206) \\ 
 Hypervigilance & 0.291 (0.223) \\ 
  Age & 0.054 (0.038) \\ 
  Domain\_Knowledge (Average) & $-$0.162 (1.035) \\ 
  Domain\_Knowledge (Below Average) & 4.173$^{***}$ (1.454) \\ 
Pursuing or Have obtained college education & $0.026$ (1.160) \\ 
    Pursuing or Have obtained graduate education & $-$0.066 (1.431) \\ 
  Perception of AI (High) & 0.885 (0.931) \\ 
  Constant & 6.950$^{**}$ (2.864) \\ 
 \hline \\[-1.8ex] 
Observations & 1,542 \\ 
Marginal R2 / Conditional R2 &	0.023 / 0.232\\
\hline 
\hline \\[-1.8ex] 
\textit{Note:}  & \multicolumn{1}{r}{$^{*}$p$<$0.1; $^{**}$p$<$0.05; $^{***}$p$<$0.01} \\ 
\end{tabular}
\Description{This image shows the results of a linear mixed model predicting individuals' time spent evaluating AI explanations. The table displays the coefficient (standard error) for each independent variable.The independent variables include Buckpassing, Vigilance, Hypervigilance, Age, Domain_Knowledge (Average), Domain_Knowledge (Below Average), Pursuing or Have obtained college education, Pursuing or Have obtained graduate education, Perception of AI (High), and Constant. The table also provides the number of observations and the Marginal R2 / Conditional R2 values. The results indicate that participants' buckpassing and vigilance tendency significantly affect the time they spent reading AI's explanations if they chose to see them. Specifically, higher buckpassing is associated with less time spent evaluating AI explanations, while higher vigilance is associated with more time spent evaluating AI explanations}
\end{table}

\begin{table}[h]
    \centering
        \caption{Ordinal Logistic Regression Model for Perceived Reliance on AI (reference levels are high school education, domain knowledge-above average, perception of AI-low). Significance levels:p<0.001***, p<0.05**, p<0.1*}
    \begin{tabular}{lcc}
        \toprule
        \multicolumn{1}{c}{\textbf{Predictors}} & \textbf{Odds Ratios} & \textbf{CI} \\
        \midrule
        Buckpassing & 1.09$^{**}$ & 1.03 -- 1.15 \\
        Hypervigilance & 0.97 & 0.91 -- 1.03 \\
        Vigilance & 0.98 & 0.93 -- 1.04 \\
        Pursuing or Have obtained college education & 1.21 & 0.89 -- 1.64 \\
        Pursuing or Have obtained graduate education & 1.60$^{*}$ & 1.08 -- 2.37 \\
        Age & 0.99 & 0.98 -- 1.00 \\
        Domain Knowledge (Average) & 1.86$^{***}$ & 1.40 -- 2.47 \\
        Domain Knowledge (Below Average) & 1.46 & 0.98 -- 2.17 \\
        Perception of AI (High) & 1.41$^{**}$ & 1.09 -- 1.82 \\
        \midrule
        Observations & \multicolumn{2}{c}{810} \\
        $R^2$ Nagelkerke & \multicolumn{2}{c}{0.056} \\
        \bottomrule
    \end{tabular}
    \Description{This image shows the results of an ordinal logistic regression model predicting perceived reliance on AI. The table displays the odds ratios, 95

The predictor variables include Buckpassing, Hypervigilance, Vigilance, Pursuing or Have obtained College Education, Pursuing or Have obtained Graduate Education, Age, Domain_Knowledge (Average), Domain_Knowledge (Below Average), and Perception of AI (High). 

The table also provides the total number of observations and the R-squared value for the Nagelkerke R-squared.

The results indicate that several factors significantly predict perceived reliance on AI, including domain knowledge, education level, and perception of AI. Specifically, higher domain knowledge (both average and below average) and higher perception of AI are associated with greater perceived reliance on AI. Additionally, pursuing or obtaining graduate education is associated with greater perceived reliance on AI compared to the reference group of high school education or less.}
\label{table:lm_reliance}

\end{table}

\subsection{Effects of Decision-making Patterns on \PerceivedAIreliance}
\begin{figure}[!h]
\centering

\includegraphics[width=0.9\linewidth]{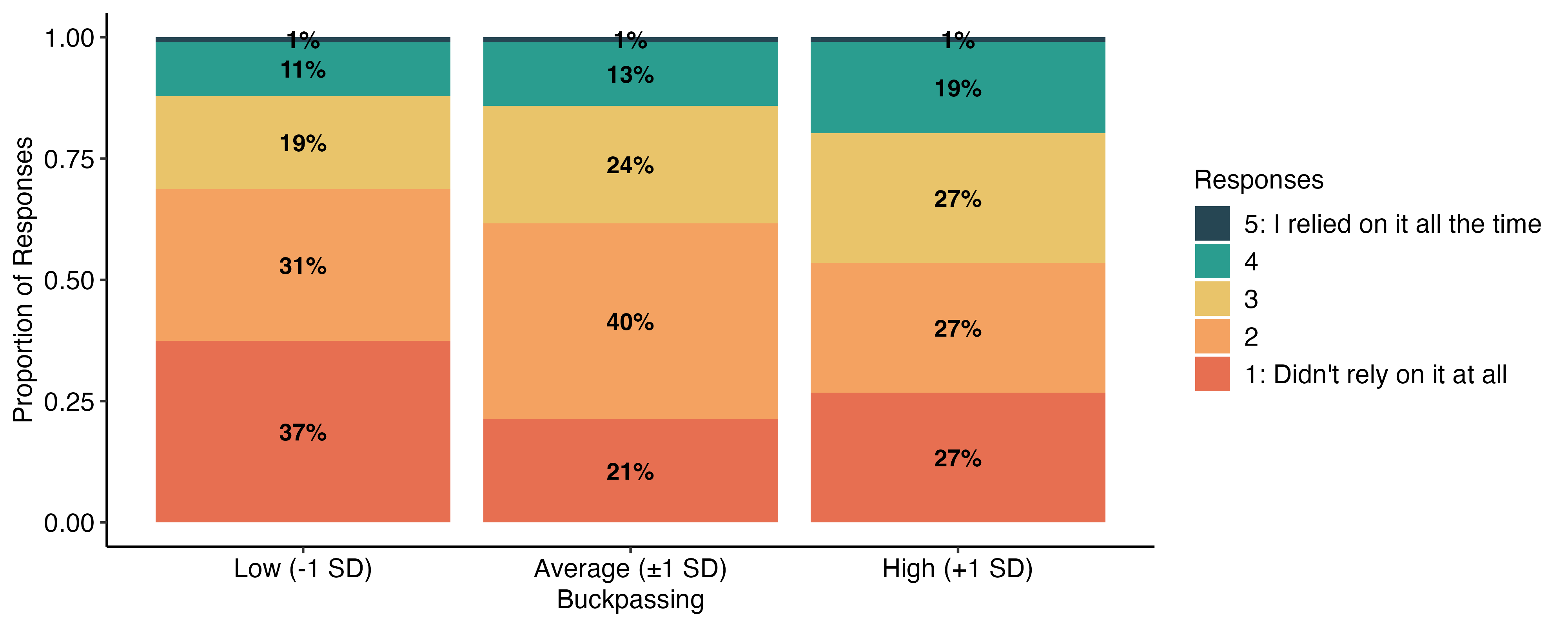}
\caption{Perceived AI Reliance Variations Across Different Levels of Buckpassing: individuals who scored high (one standard deviation above average among participants)  in buckpassing reported a higher reliance on AI compared to those who scored low (one standard deviation below average) in buckpassing.}
\label{buckpassing_reliance}
\Description{This image shows the perceived AI reliance variations across different levels of buckpassing. The x-axis represents the three levels of buckpassing: low (-1 SD), average (+1 SD), and high (+1 SD). The y-axis shows the proportion of responses for each level of perceived reliance on AI, ranging from "1: Didn't rely on it at all" to "5: I relied on it all the time".

The key findings are:
Individuals who score high in buckpassing tend to report a higher reliance on AI. For the high buckpassing group, 19\% selected "4" or "5" on the perceived reliance scale, compared to 13\% for the average buckpassing group and 11\% for the low buckpassing group.
Conversely, those with low buckpassing are more likely to select "1: Didn't rely on it at all", with 37\% of the low buckpassing group choosing this response, compared to 21\% for the average group and 27\% for the high buckpassing group.

The overall pattern suggests that individuals with a higher tendency to engage in buckpassing behaviors are more likely to report a greater perceived reliance on AI systems.}
\end{figure}

To test \textbf{H1c}, \textbf{H2c}, and \textbf{H3c}, we constructed an ordinal logistic regression model while controlling for participants' demographic covariates including age, education level, domain knowledge, and perception of AI. Ordinal logistic regression is often used to predict dependent variables that can be ordered in a natural way such as \textit{mild, moderate, severe} \cite{Harrell2015}. Per \autoref{table:lm_reliance}, our result indicates the effect of buckpassing on perceived AI reliance was statistically significant ($OR = 1.09, 95\% \text{ CI} [1.03, 1.15], p<0.05$). For one score of increase in buckpassing, the odds of having more self-perceived reliance on the AI increases by 9\%, suggesting that people who score high in buckpassing are more likely to rely on AI when making their decisions. Figure \ref{buckpassing_reliance} shows the distribution of response in terms of level of reliance related to individuals' buckpassing tendency. Noteworthily, 12\% of people who received a low buckpassing score reported to rely on the AI ``all the time'' (rating of 5) or most of the time (rating of 4), while 20\% of those who received a high buckpassing score reported the same. 

No effect of vigilance or hypervigilance was observed. Hypotheses \textbf{H1c} and \textbf{H2c} are therefore not supported, while \textbf{H3c} is supported.

\subsection{Who tends to score high in buckpassing?}

\begin{figure}
    \centering
    \includegraphics[width=\linewidth]{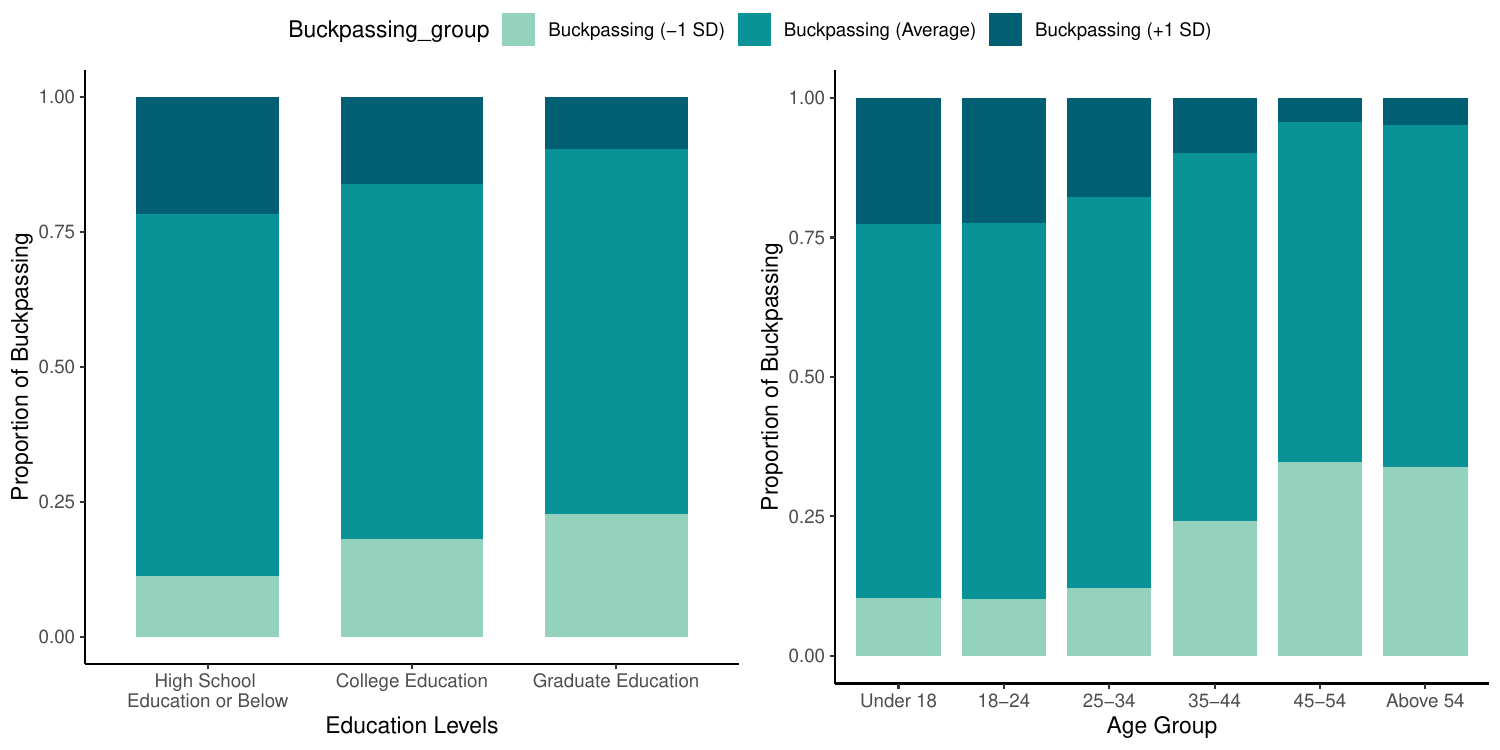}
 \caption{Examining the demographic distribution of buckpassing tendency: participants who have a high school education or below score higher in buckpassing (23\% in high buckpassing level); participants who are aged under 18 and in their early 20s also tend to score higher in buckpassing. The y-axis is the proportion of participants in each buckpassing group.}
    \label{fig:buckpassing_distribution}
\end{figure}
Our analysis indicates that a buckpassing tendency significantly influences participants' use of AI suggestions and reliance on AI. 
Thus we further explored the demographic distribution of buckpassing scores among our participants to gain insights into which demographic groups exhibit a higher buckpassing tendency.  

Figure \ref{fig:buckpassing_distribution} shows the distribution of buckpassing levels among different education levels and age groups. Age is treated as a categorical variable in this descriptive analysis to obtain a more granular understanding of age group differences.  Teenagers and young adults in our sample tended to score higher in buckpassing compared to participants who were aged above 35. Participants who obtain a lower level of education (high school education or below) also exhibited a higher buckpassing tendency compared to those with a higher level of education (pursuing or having obtained college or graduate education). 

\section{Discussion}
Individuals’ decision-making patterns have been shown to influence their  decision outcomes and quality of life~\cite{kim_nowhere_2022,deniz2006relationships}, underlining the importance of understanding how people vary in making decisions. However, despite the increasing role of AI in decision-making contexts, how individual cognitive decision-making processes influence the use of AI technology remains unknown. Our study offers evidence of this influence; we discuss our findings and consider their implications for designing AI systems and explainable AI research below. 
\paragraph{Decision-making Patterns as Important Factors in AI-assisted Decision-making}
In this work, we extend the current understanding of individuals' interactions with AI-generated information and factors behind these interactions in a setting where AI assistance is available but optional. Consistent with our hypotheses, we found that people who tend to defer decision-making patterns to others (score high in buckpassing) are more likely to seek out suggestions from an AI and report relying on it, though they spend less time reading AI’s explanations. While the difference can be described as a small to medium effect---those who score low on buckpassing viewed the AI's decisions 25\% of the time versus 33\% for those who scored high on buckpassing---the effect of this difference can be profound when AI decision-making tools are frequently used and the performance of AI tools is not high. As we observed in our data, participants scored lower in accuracy when exposed to inaccurate AI decisions and higher when exposed to correct AI decisions. The difference may also gain importance with the number of decisions to be made as decision fatigue sets in~\cite{10.1145/3610219} or as task difficulty or uncertainty increases~\cite{vasconcelos_explanations_2023,salimzadeh2024dealing}, leading to a higher likelihood of overreliance. 


We encourage future work to examine these factors in various kinds of human-AI interactions. 
In addition, prior HCI research shows that individuals' trust and reliance vary by their age, education levels, and perception of AI~\cite{MA2023102362,araujo_ai_2020}. Our work extends this by showing that individuals’ buckpassing tendency has a marginal effect on AI use when accounting for these demographic covariates. While the significance of the effect decreases after controlling these factors, this is likely due to the correlation between these demographic factors and decision-making patterns (see  Appendix \autoref{correlation_matrix} for the correlation matrix).

What might explain these findings is the stress level and confidence people experience while making decisions. \citet{mann_melbourne_1997}'s work has indicated that these decision-making tendencies can be seen as different kinds of coping mechanisms during decision conflict. Individuals with a strong buckpassing tendency often experience high psychological stress during decision-making tasks, while vigilant decision-makers experience a lower level of stress~\cite{mann_melbourne_1997}. Experiencing high psychological stress while making decisions might propel people to seek out an immediate relief by turning to AI suggestions.  \citet{mann_cross-cultural_1998}
also identified that individuals who score high in buckpassing tend to have lower confidence in their own decision-making. This could potentially propel participants in our study to seek AI's suggestions when they were not confident about making the decisions themselves. 

Interestingly, we did not observe any effect of vigilance or hypervigilance on participants' tendency to see AI suggestions.  This might be due to the low variance of participants on vigilance 
 ($SD = 2.31$) and hypervigilance ($SD = 2.61$) compared to buckpassing pattern ($3.03$).  However, we observed that participants who scored high in vigilance spent significantly more time seeing AI explanations. This suggests that increased vigilance goes hand in hand with careful reading of any explanations people are provided, likely as a tool for evaluating whether the AI's decision can be reasonably trusted.

\paragraph{Benefits and Risks of AI-assisted Decision-making for People with Different Decision-Making Patterns}
AI-assisted decision-making could yield both positive and negative outcomes for people with a tendency for buckpassing. On the one hand, the rise of AI systems could ease the burden of information seeking and processing for individuals to make effective decisions, if the performance of these systems is reliable. As evidenced by prior work, making decisions with the help of AI could lead to more accurate decisions when the AI has high advice accuracy and quality~\cite{vodrahalli_humans_2021,tschandl2020human} (though there are some exceptions where this was found not to be the case~\cite{vaccaro2019effects,jacobs2021machine}). AI assistance can also lower mental demand~\cite{bucinca_trust_2021}. %
With highly reliable AI systems, AI suggestions could help relieve the decision-making stress experienced by people with a buckpassing tendency. 
On the other hand, given the current risks of AI hallucinations, where AI generates unfaithful or nonfactual content~\cite{bubeck2023sparks}, and AI overconfidence~\cite{zhou-etal-2024-relying}, where AI outputs sound confident yet are inaccurate~\cite{zhou-etal-2024-relying}, buckpassers are more susceptible to inaccurate information generated by AI, potentially resulting in misleading decisions and negative decision outcomes.  

The benefits and risks of AI-assisted decision-making might be exacerbated for some people and groups, as our study and other research show that decision-making patterns vary across demographics. Consistent with prior findings~\cite{filipe_validation_2020}, our study suggests that buckpassing tendency is more prevalent among younger people (especially individuals under 18) and people with high school education or below. This warrants greater intervention to guide these populations in their usage of AI systems. Prior work from \citet{mann_cross-cultural_1998} indicates that people from East Asian countries  (e.g., Japan, Hong Kong) exhibit a higher buckpassing tendency compared to people in Western countries (e.g., USA, Australia, New Zealand). East Asians have also been found to have a lower decision-making self-esteem, which is generally in line with a preference for making decisions in a group~\cite{mann_cross-cultural_1998}.  These demographic groups may seek out information from the AI more often, benefiting more from AI systems that have high accuracy but also facing greater risks of using inaccurate AI-generated information.

\subsection{Implications for Designing AI Systems and Explainable AI Research}

The variations in people's decision-making patterns, and in particular in people's tendency to be vigilant or pass the buck warrant more caution in the design and implementation of AI systems. How do we cultivate appropriate reliance on AI systems given these decision-making patterns? Prior work has investigated several interventions that could help mitigate inappropriate reliance, which we consider together with the implications of our findings below. 

\paragraph{Account for the differential effects of cognitive forcing functions} Our findings showed that those prone to deferring decisions to others are also more likely to seek out AI suggestions. The relatively short time they spent looking at the AI explanations, compared to participants who scored low in buckpassing, further suggests that buckpassers may not engage as strongly with this information; the immediate access to the AI suggestions (as is most common in human-AI studies and AI decision-support systems in various contexts~\cite{10.1145/3610219,morrison_evaluating_2023,10.1145/3411764.3445522,yin2019understanding,cao_how_2023,lu_does_2024}) seems to be an easy way to lower cognitive effort, which may lead to overreliance on the AI. Cognitive forcing functions---like presenting AI suggestions on demand or forcing people to make their own decisions \emph{before} seeing an AI's suggestions---are effective strategies researchers have identified in reducing overreliance~\cite{bucinca_proxy_2020}. Yet it remains unclear how individuals differ in interacting with each kind of cognitive forcing function. Our results suggest that decision-making patterns also influence individuals' engagement with AI suggestions when they are presented on demand. In addition, since buckpassers often defer decision-making to others due to experienced stress, implementing cognitive forcing functions may lead to more mental demand and stress in decision-making. Thus, designers should prioritize relatively low-effort cognitive forcing functions and couple them with motivational nudges~\cite{caraban2019nudge,vasconcelos_explanations_2023} (e.g., by reminding the potential consequences of overreliance behaviors~\cite{bo2025to}) to encourage active engagement among this group. 
%




\paragraph{Adopt presentation formats of AI explanations that ease cognitive load and decision stress.} Our study found that buckpassers tend to spend less, and vigilant people more time, when  reviewing AI explanations, posing challenges to the design of AI explanations. The format of AI explanations in our study was text-based and lengthy. This format might induce further cognitive load and stress for buckpassers, resulting in their reduced time investment in AI explanations. Prior research found that if decisions are made under high cognitive load, they can lead to higher overreliance on AI~\cite{Zhang2024}. Thus, similar to cognitive forcing functions, identifying presentation formats of AI suggestions that reduce cognitive load and decision stress may encourage buckpassers to evaluate AI explanations carefully.
\citet{10.1145/3610219} find that for individuals who rely on their intuition in decision-making, providing example-based scenarios as AI explanations can effectively calibrate appropriate reliance on AI. 
Psychology research on cognitive load theory (CLT) and e-learning has identified various information presentation techniques to reduce individuals' cognitive load~\cite{KIRSCHNER20021,van2005research,sweller2011cognitive}. \citet{mousavi1995reducing} found that presenting information in both audio and visual format leads to less cognitive load than in visual format alone. As such, designers could consider showing AI explanations to users in both visual and audio formats to lower their cognitive load. Redundant information is also shown to increase individuals' cognitive load~\cite{sweller2011cognitive}. However, experts and novices have different perceptions of redundancy; in AI suggestions, designers should personalize information based on individuals' domain knowledge  and avoid providing information that is commonly known among domain experts.

\paragraph{Provide users with diverse options for the presentation type of AI suggestions}
Despite the effectiveness of interventions in cultivating appropriate reliance on AI, research in explainable AI has focused on explanations for binary decision-making tasks; however,  AI systems often generate information in a limited presentation format and a lack of user control. The varied amount of time investment could result in a different level of caution when processing information from AI. While these lengthy paragraphs might suit the information-seeking habits of vigilant people, they might not be processed carefully by individuals who tend to be buckpassers. As such, developers should consider providing features that enable users to select their own preferred presentation format based on their own information-seeking habits. For example, if users may adjust the length of text and amount of details (i.e. summary of references or concrete examples) used in AI suggestions, vigilant individuals may choose to obtain more comprehensive information whereas buckpassers could request brief explanations.

\section{Limitations and Future Work}
We acknowledge several limitations of this study. First, the observed behaviors in the design of this study may not generalize to other domains and tasks. Our study was designed to capture participants' collaborative decision-making behavior with an AI in a task that required them to accurately identify whether a set of nutrition statements are facts or myths. Choosing nutrition as a domain has the advantage that the study was broadly accessible to diverse participants, but inaccuracies have naturally less severe consequences than tasks in, e.g., the medical domain.  We additionally chose to conduct an online study with volunteer participants who were offered personalized performance feedback in exchange for study completion. This usually means participants are intrinsically motivated to do well; however, the consequences of mistaking a nutrition statement as true when it was false (and vice versa) are mild when compared to a real-world situation. While similar effects of decision-making patterns may also occur in other domains and general-purpose tasks, these effects could be influenced by task differences. For tasks that are more complex, these effects may be exacerbated by the complexity of the task~\cite{bansal2019beyond}. For tasks that have less objective ground truth, these effects may still persist yet decision outcomes may change based on individuals' expertise. Future work is needed to confirm how task complexity and outcome severity shape these effects. We encourage researchers to adopt similar study designs in other settings and professional domains to better understand these dynamics.

Secondly, our study did not fully capture individuals' various reliance outcomes. Prior studies on AI-assisted decision-making investigated individuals' appropriate reliance (overreliance or underreliance) on AI~\cite{cao_designing_2024,vasconcelos_explanations_2023}. These measures were obtained in a study design that involves two-stage decision-making (e.g., individuals are asked to make decisions first before they are exposed to AI suggestions, and then make final decisions after seeing AI suggestions). Our study adopted a design setting different from two-stage decision-making, because we focused on whether participants seek out AI suggestions and how they perceive their reliance on the AI overall. This means that our study outcome is only partially comparable to prior findings on AI overreliance. Specifically, while we observed that participants who scored high in buckpassing tend to perform worse than those who scored low in buckpassing when both were exposed to incorrect AI decisions, this difference was not significant. Therefore, while participants with high buckpassing tendency were significantly more likely to seek out AI suggestions, we cannot draw firm conclusions about how this tendency influences their overall decision performance.  Future work is needed to disentangle whether buckpassers tend to overrely on AI in addition to being more likely to seek AI assistance.

Thirdly, our study design did not fully capture users' holistic experiences with AI systems, such as ChatGPT, through chatbot interfaces since users did not have a choice to see more AI explanations if they were not satisfied with the output. We chose such a design to ensure AI suggestions are consistent for each participant, yet we hope future studies could explore experiment designs that closely mimic interactions with generative AI systems. Meanwhile, our design of asking users' to indicate their adoption of AI suggestion by providing an option ``Use the AI Decision'' had very few engagement click despite the high reliance suggested by other measures. We suspect this occurred due to an unintuitive display of options for users to choose from or a motivation to preserve autonomy in making the final decision. In future work, we will also improve our designs of interface features that are more closely matched with users' behaviors. It's also worth noting that our design of the study does not provide the opportunity to search for information from the web. In real-world scenarios, users could potentially navigate other information sources for more information before they make their final decisions. Future work could further integrate this option to examine how users' behaviors are influenced by their decision-making patterns in a more dynamic setting. Notably, while we did not examine the effect of procrastination in our study, future work that integrates more dynamic interactions shall investigate the behaviors of individuals who tend to procrastinate in their decision-making. 

Lastly, our sample primarily consisted of participants who were fluent in English and motivated to participate in our study due to our study topic.   
In future work, we hope to gather a larger and more diverse sample via by providing the study in other languages and various topics to follow up on the present findings. 
In line with this, one exciting avenue for future work is to investigate how differences in decision-making patterns across countries and cultures impact people's reliance on an AI. For instance, \citet{mann_cross-cultural_1998} had found that people from more hierarchical cultures commonly score high in buckpassing, which could imply that there are cross-cultural differences in reliance, and potentially in overreliance, on an AI. 

\section{Conclusion}
With the increasing use of conversational AI such as ChatGPT for information-seeking and decision-making, it is essential to understand how people vary in their interactions with AI suggestions in the decision-making process.  Through an online study, we asked participants (n=810) to evaluate the factuality of nutrition-related statements with the option to seek AI suggestions (decisions and explanations). We found that people who tend to defer decisions to others (buckpassers) are more likely to seek AI suggestions yet spend less time evaluating these suggestions and reported a higher level of reliance on AI  when evaluating nutrition information than those scoring low on buckpassing. In contrast, vigilant decision makers tend to more carefully scrutinize the AI's information than those scoring low on vigilance. Drawing insight from psychology research on decision-making, our study suggests that individuals’ decision-making patterns implicate not only \textit{human-human} interactions but also human-AI interactions.  In particular, these findings expand the current research horizon of AI-assisted decision-making by underscoring the importance of individual cognitive processes. As more AI-driven systems are being developed and integrated into our everyday lives, these findings shed light on the importance of individual cognitive factors, providing new insights for the future development of AI technologies.


\begin{acks}
    This work is funded by the National Science Foundation under Grant No.2230466.
\end{acks}

\bibliographystyle{ACM-Reference-Format}
\bibliography{10_reference} 

@article{cacioppo1982need,
  title={The need for cognition.},
  author={Cacioppo, John T and Petty, Richard E},
  journal={Journal of personality and social psychology},
  volume={42},
  number={1},
  pages={116},
  year={1982},
  publisher={American Psychological Association}
}

@article{lee2004trust,
  title={Trust in automation: Designing for appropriate reliance},
  author={Lee, John D and See, Katrina A},
  journal={Human factors},
  volume={46},
  number={1},
  pages={50--80},
  year={2004},
  publisher={SAGE Publications Sage UK: London, England}
}

@article{papenmeier2022s,
  title={It’s complicated: The relationship between user trust, model accuracy and explanations in AI},
  author={Papenmeier, Andrea and Kern, Dagmar and Englebienne, Gwenn and Seifert, Christin},
  journal={ACM Transactions on Computer-Human Interaction (TOCHI)},
  volume={29},
  number={4},
  pages={1--33},
  year={2022},
  publisher={ACM New York, NY}
}

@article{Zhang2024,
  title = {Measuring the effect of mental workload and explanations on appropriate AI reliance using EEG},
  ISSN = {1362-3001},
  url = {http://dx.doi.org/10.1080/0144929X.2024.2431055},
  DOI = {10.1080/0144929x.2024.2431055},
  journal = {Behaviour  \&  Information Technology},
  publisher = {Informa UK Limited},
  author = {Zhang,  Zelun Tony and Argın,  Seniha Ketenci and Bilen,  Mustafa Baha and Urgun,  Doğan and Deniz,  Sencer Melih and Liu,  Yuanting and Hassib,  Mariam},
  year = {2024},
  month = nov,
  pages = {1–19}
}

@article{tschandl2020human,
  title={Human--computer collaboration for skin cancer recognition},
  author={Tschandl, Philipp and Rinner, Christoph and Apalla, Zoe and Argenziano, Giuseppe and Codella, Noel and Halpern, Allan and Janda, Monika and Lallas, Aimilios and Longo, Caterina and Malvehy, Josep and others},
  journal={Nature medicine},
  volume={26},
  number={8},
  pages={1229--1234},
  year={2020},
  publisher={Nature Publishing Group US New York}
}

@inproceedings{salimzadeh2024dealing,
  title={Dealing with Uncertainty: Understanding the Impact of Prognostic Versus Diagnostic Tasks on Trust and Reliance in Human-AI Decision Making},
  author={Salimzadeh, Sara and He, Gaole and Gadiraju, Ujwal},
  booktitle={Proceedings of the CHI Conference on Human Factors in Computing Systems},
  pages={1--17},
  year={2024}
}

@article{jacobs2021machine,
  title={How machine-learning recommendations influence clinician treatment selections: the example of antidepressant selection},
  author={Jacobs, Maia and Pradier, Melanie F and McCoy Jr, Thomas H and Perlis, Roy H and Doshi-Velez, Finale and Gajos, Krzysztof Z},
  journal={Translational psychiatry},
  volume={11},
  number={1},
  pages={108},
  year={2021},
  publisher={Nature Publishing Group UK London}
}

@article{vaccaro2019effects,
  title={The effects of mixing machine learning and human judgment},
  author={Vaccaro, Michelle and Waldo, Jim},
  journal={Communications of the ACM},
  volume={62},
  number={11},
  pages={104--110},
  year={2019},
  publisher={ACM New York, NY, USA}
}

@article{rozado2023political,
  title={The political biases of chatgpt},
  author={Rozado, David},
  journal={Social Sciences},
  volume={12},
  number={3},
  pages={148},
  year={2023},
  publisher={MDPI}
}

@article{lu_does_2024,
	title = {Does {More} {Advice} {Help}? {The} {Effects} of {Second} {Opinions} in {AI}-{Assisted} {Decision} {Making}},
	volume = {8},
	issn = {2573-0142},
	shorttitle = {Does {More} {Advice} {Help}?},
	url = {https://dl.acm.org/doi/10.1145/3653708},
	doi = {10.1145/3653708},
	language = {en},
	number = {CSCW1},
	urldate = {2024-05-23},
	journal = {Proceedings of the ACM on Human-Computer Interaction},
	author = {Lu, Zhuoran and Wang, Dakuo and Yin, Ming},
	month = apr,
	year = {2024},
	pages = {1--31},
}

@article{chowdhury_time-harried_2009,
	title = {The time-harried shopper: {Exploring} the differences between maximizers and satisficers},
	volume = {20},
	issn = {1573-059X},
	shorttitle = {The time-harried shopper},
	url = {https://doi.org/10.1007/s11002-008-9063-0},
	doi = {10.1007/s11002-008-9063-0},
	language = {en},
	number = {2},
	urldate = {2024-07-05},
	journal = {Marketing Letters},
	author = {Chowdhury, Tilottama G. and Ratneshwar, S. and Mohanty, Praggyan},
	month = jun,
	year = {2009},
	pages = {155--167},
}

@article{simon_behavioral_1955,
	title = {A {Behavioral} {Model} of {Rational} {Choice}},
	volume = {69},
	issn = {0033-5533},
	url = {https://www.jstor.org/stable/1884852},
	doi = {10.2307/1884852},
	number = {1},
	urldate = {2024-07-05},
	journal = {The Quarterly Journal of Economics},
	author = {Simon, Herbert A.},
	year = {1955},
	note = {Publisher: Oxford University Press},
	pages = {99--118},
}

@article{cao_designing_2024,
	title = {Designing for {Appropriate} {Reliance}: {The} {Roles} of {AI} {Uncertainty} {Presentation}, {Initial} {User} {Decision}, and {User} {Demographics} in {AI}-{Assisted} {Decision}-{Making}},
	volume = {8},
	issn = {2573-0142},
	shorttitle = {Designing for {Appropriate} {Reliance}},
	url = {https://dl.acm.org/doi/10.1145/3637318},
	doi = {10.1145/3637318},
	language = {en},
	number = {CSCW1},
	urldate = {2024-05-09},
	journal = {Proceedings of the ACM on Human-Computer Interaction},
	author = {Cao, Shiye and Liu, Anqi and Huang, Chien-Ming},
	month = apr,
	year = {2024},
	pages = {1--32},
}

@article{kapelner2010preventing,
  title={Preventing satisficing in online surveys},
  author={Kapelner, Adam and Chandler, Dana},
  journal={Proceedings of CrowdConf},
  year={2010}
}

@inproceedings{kim_humans_2023,
	address = {New York, NY, USA},
	series = {{FAccT} '23},
	title = {Humans, {AI}, and {Context}: {Understanding} {End}-{Users}’ {Trust} in a {Real}-{World} {Computer} {Vision} {Application}},
	isbn = {9798400701924},
	shorttitle = {Humans, {AI}, and {Context}},
	url = {https://dl.acm.org/doi/10.1145/3593013.3593978},
	doi = {10.1145/3593013.3593978},
	urldate = {2023-06-19},
	booktitle = {Proceedings of the 2023 {ACM} {Conference} on {Fairness}, {Accountability}, and {Transparency}},
	publisher = {Association for Computing Machinery},
	author = {Kim, Sunnie S. Y. and Watkins, Elizabeth Anne and Russakovsky, Olga and Fong, Ruth and Monroy-Hernández, Andrés},
	month = jun,
	year = {2023},
	pages = {77--88},
}

@inproceedings{lai_towards_2023,
	address = {New York, NY, USA},
	series = {{FAccT} '23},
	title = {Towards a {Science} of {Human}-{AI} {Decision} {Making}: {An} {Overview} of {Design} {Space} in {Empirical} {Human}-{Subject} {Studies}},
	isbn = {9798400701924},
	shorttitle = {Towards a {Science} of {Human}-{AI} {Decision} {Making}},
	url = {https://dl.acm.org/doi/10.1145/3593013.3594087},
	doi = {10.1145/3593013.3594087},
	urldate = {2023-06-19},
	booktitle = {Proceedings of the 2023 {ACM} {Conference} on {Fairness}, {Accountability}, and {Transparency}},
	publisher = {Association for Computing Machinery},
	author = {Lai, Vivian and Chen, Chacha and Smith-Renner, Alison and Liao, Q. Vera and Tan, Chenhao},
	month = jun,
	year = {2023},
	pages = {1369--1385},
}

@article{mann_cross-cultural_1998,
	title = {Cross-cultural {Differences} in {Self}-reported {Decision}-making {Style} and {Confidence}},
	volume = {33},
	issn = {1464-066X},
	url = {https://onlinelibrary.wiley.com/doi/abs/10.1080/002075998400213},
	doi = {10.1080/002075998400213},
	language = {fr},
	number = {5},
	urldate = {2023-06-20},
	journal = {International Journal of Psychology},
	author = {Mann, Leon and Radford, Mark and Burnett, Paul and Ford, Steve and Bond, Michael and Leung, Kwok and Nakamura, Hiyoshi and Vaughan, Graham and Yang, Kuo-Shu},
	year = {1998},
	pages = {325--335},
}

@inproceedings{bucinca_proxy_2020,
	title = {Proxy {Tasks} and {Subjective} {Measures} {Can} {Be} {Misleading} in {Evaluating} {Explainable} {AI} {Systems}},
	url = {http://arxiv.org/abs/2001.08298},
	doi = {10.1145/3377325.3377498},
	language = {en},
	urldate = {2023-06-20},
	booktitle = {Proceedings of the 25th {International} {Conference} on {Intelligent} {User} {Interfaces}},
	author = {Buçinca, Zana and Lin, Phoebe and Gajos, Krzysztof Z. and Glassman, Elena L.},
	month = mar,
	year = {2020},
	note = {arXiv:2001.08298 [cs]},
	pages = {454--464},
}

@article{de2004decision,
  title={Decision-making patterns, conflict sytles, and self-esteem},
  author={De Heredia, Ram{\'o}n Alzate S{\'a}ez and Arocena, Francisco Laca and G{\'a}rate, Jos{\'e} Valencia},
  journal={Psicothema},
  pages={110--116},
  year={2004}
}

@article{alexander2017reported,
  title={Reported maladaptive decision-making in unipolar and bipolar depression and its change with treatment},
  author={Alexander, Lara F and Oliver, Alison and Burdine, Lauren K and Tang, Yilang and Dunlop, Boadie W},
  journal={Psychiatry research},
  volume={257},
  pages={386--392},
  year={2017},
  publisher={Elsevier}
}

@article{bouckenooghe2007cognitive,
  title={Cognitive motivation correlates of coping style in decisional conflict},
  author={Bouckenooghe, Dave and Vanderheyden, Karlien and Mestdagh, Steven and Van Laethem, Sarah},
  journal={The Journal of Psychology},
  volume={141},
  number={6},
  pages={605},
  year={2007},
  publisher={Taylor \& Francis Inc.}
}

@article{brown2016decision,
  title={Decision-making approaches and the propensity to default: Evidence and implications},
  author={Brown, Jeffrey R and Farrell, Anne M and Weisbenner, Scott J},
  journal={Journal of Financial Economics},
  volume={121},
  number={3},
  pages={477--495},
  year={2016},
  publisher={Elsevier}
}

@article{nota2000adattamento,
  title={Adattamento italiano del melbourne decision making questionnaire di leon mann},
  author={Nota, Laura and Soresi, Salvatore and others},
  journal={GIPO-GIORNALE ITALIANO DI PSICOLOGIA DELL'ORIENTAMENTO},
  volume={3},
  pages={38--52},
  year={2000},
  publisher={Giunti-OS}
}

@article{colakkadioglu2015study,
  title={Study on the validity and reliability of Melbourne Decision Making Scale in Turkey},
  author={Colakkadioglu, Oguzhan and Deniz, M Engin},
  journal={Educational Research and Reviews},
  volume={10},
  number={10},
  pages={1434--1441},
  year={2015},
  publisher={Academic Journals}
}

@article{filipe_validation_2020,
	title = {Validation and invariance across age and gender for the {Melbourne} {Decision}-{Making} {Questionnaire} in a sample of {Portuguese} adults},
	volume = {15},
	issn = {1930-2975},
	url = {https://www.cambridge.org/core/journals/judgment-and-decision-making/article/validation-and-invariance-across-age-and-gender-for-the-melbourne-decisionmaking-questionnaire-in-a-sample-of-portuguese-adults/09643F44F9D05F4CDF85EF2B1FBEFDA4},
	doi = {10.1017/S1930297500006951},
	language = {en},
	number = {1},
	urldate = {2023-12-21},
	journal = {Judgment and Decision Making},
	author = {Filipe, Luís and Alvarez, Maria-João and Roberto, Magda Sofia and Ferreira, Joaquim A.},
	month = jan,
	year = {2020},
	note = {Publisher: Cambridge University Press},
	pages = {135--148},
}

@article{bailly2011adaptation,
  title={Adaptation et validation en langue Fran{\c{c}}aise d'une {\'e}chelle de prise de d{\'e}cision.},
  author={Bailly, Nathalie and Ilharragorry-Devaux, Marie-Lise},
  journal={Canadian Journal of Behavioural Science/Revue canadienne des sciences du comportement},
  volume={43},
  number={3},
  pages={143},
  year={2011},
  publisher={Educational Publishing Foundation}
}

@article{Cotrena2017AdaptationAV,
  title={Adaptation and validation of the Melbourne Decision Making Questionnaire to Brazilian Portuguese.},
  author={Charles Cotrena and Laura Damiani Branco and Rochele Paz Fonseca},
  journal={Trends in psychiatry and psychotherapy},
  year={2017},
  volume={40 1},
  pages={
          29-37
        },
  url={https://api.semanticscholar.org/CorpusID:4971830}
}

@article{bucinca_trust_2021,
	title = {To {Trust} or to {Think}: {Cognitive} {Forcing} {Functions} {Can} {Reduce} {Overreliance} on {AI} in {AI}-assisted {Decision}-making},
	volume = {5},
	issn = {2573-0142},
	shorttitle = {To {Trust} or to {Think}},
	url = {http://arxiv.org/abs/2102.09692},
	doi = {10.1145/3449287},
	number = {CSCW1},
	urldate = {2023-06-21},
	journal = {Proceedings of the ACM on Human-Computer Interaction},
	author = {Buçinca, Zana and Malaya, Maja Barbara and Gajos, Krzysztof Z.},
	month = apr,
	year = {2021},
	note = {arXiv:2102.09692 [cs]},
	pages = {1--21},
}

@article{guerreiro2023hallucinations,
  title={Hallucinations in large multilingual translation models},
  author={Guerreiro, Nuno M and Alves, Duarte M and Waldendorf, Jonas and Haddow, Barry and Birch, Alexandra and Colombo, Pierre and Martins, Andr{\'e} FT},
  journal={Transactions of the Association for Computational Linguistics},
  volume={11},
  pages={1500--1517},
  year={2023},
  publisher={MIT Press One Broadway, 12th Floor, Cambridge, Massachusetts 02142, USA~…}
}

@inproceedings{zhang_effect_2020,
	address = {Barcelona Spain},
	title = {Effect of confidence and explanation on accuracy and trust calibration in {AI}-assisted decision making},
	isbn = {978-1-4503-6936-7},
	url = {https://dl.acm.org/doi/10.1145/3351095.3372852},
	doi = {10.1145/3351095.3372852},
	language = {en},
	urldate = {2023-06-22},
	booktitle = {Proceedings of the 2020 {Conference} on {Fairness}, {Accountability}, and {Transparency}},
	publisher = {ACM},
	author = {Zhang, Yunfeng and Liao, Q. Vera and Bellamy, Rachel K. E.},
	month = jan,
	year = {2020},
	pages = {295--305},
}

@inproceedings{lai_human_2019,
	address = {New York, NY, USA},
	series = {{FAT}* '19},
	title = {On {Human} {Predictions} with {Explanations} and {Predictions} of {Machine} {Learning} {Models}: {A} {Case} {Study} on {Deception} {Detection}},
	isbn = {978-1-4503-6125-5},
	shorttitle = {On {Human} {Predictions} with {Explanations} and {Predictions} of {Machine} {Learning} {Models}},
	url = {https://dl.acm.org/doi/10.1145/3287560.3287590},
	doi = {10.1145/3287560.3287590},
	urldate = {2023-06-23},
	booktitle = {Proceedings of the {Conference} on {Fairness}, {Accountability}, and {Transparency}},
	publisher = {Association for Computing Machinery},
	author = {Lai, Vivian and Tan, Chenhao},
	month = jan,
	year = {2019},
	pages = {29--38},
}

@article{cao_understanding_2022,
	title = {Understanding {User} {Reliance} on {AI} in {Assisted} {Decision}-{Making}},
	volume = {6},
	issn = {2573-0142},
	url = {https://dl.acm.org/doi/10.1145/3555572},
	doi = {10.1145/3555572},
	language = {en},
	number = {CSCW2},
	urldate = {2023-06-23},
	journal = {Proceedings of the ACM on Human-Computer Interaction},
	author = {Cao, Shiye and Huang, Chien-Ming},
	month = nov,
	year = {2022},
	pages = {1--23},
}

@inproceedings{wang_are_2021,
	address = {College Station TX USA},
	title = {Are {Explanations} {Helpful}? {A} {Comparative} {Study} of the {Effects} of {Explanations} in {AI}-{Assisted} {Decision}-{Making}},
	isbn = {978-1-4503-8017-1},
	shorttitle = {Are {Explanations} {Helpful}?},
	url = {https://dl.acm.org/doi/10.1145/3397481.3450650},
	doi = {10.1145/3397481.3450650},
	language = {en},
	urldate = {2023-06-27},
	booktitle = {26th {International} {Conference} on {Intelligent} {User} {Interfaces}},
	publisher = {ACM},
	author = {Wang, Xinru and Yin, Ming},
	month = apr,
	year = {2021},
	pages = {318--328},
}

@article{cao_how_2023,
	title = {How {Time} {Pressure} in {Different} {Phases} of {Decision}-{Making} {Influences} {Human}-{AI} {Collaboration}},
	volume = {7},
	issn = {2573-0142},
	url = {https://dl.acm.org/doi/10.1145/3610068},
	doi = {10.1145/3610068},
	language = {en},
	number = {CSCW2},
	urldate = {2024-05-14},
	journal = {Proceedings of the ACM on Human-Computer Interaction},
	author = {Cao, Shiye and Gomez, Catalina and Huang, Chien-Ming},
	month = sep,
	year = {2023},
	pages = {1--26},
}

@inproceedings{jugovac2018investigating,
  title={Investigating the decision-making behavior of maximizers and satisficers in the presence of recommendations},
  author={Jugovac, Michael and Nunes, Ingrid and Jannach, Dietmar},
  booktitle={Proceedings of the 26th Conference on User Modeling, Adaptation and Personalization},
  pages={279--283},
  year={2018}
}

@Inbook{Harrell2015,
author="Harrell, Frank E.",
title="Ordinal Logistic Regression",
bookTitle="Regression Modeling Strategies: With Applications to Linear Models, Logistic and Ordinal Regression, and Survival Analysis",
year="2015",
publisher="Springer International Publishing",
address="Cham",
pages="311--325",
isbn="978-3-319-19425-7",
doi="10.1007/978-3-319-19425-7_13",
url="https://doi.org/10.1007/978-3-319-19425-7_13"
}

@article{tavakol_making_2011,
	title = {Making sense of {Cronbach}'s alpha},
	volume = {2},
	issn = {2042-6372},
	url = {https://www.ncbi.nlm.nih.gov/pmc/articles/PMC4205511/},
	doi = {10.5116/ijme.4dfb.8dfd},
	urldate = {2024-06-19},
	journal = {International Journal of Medical Education},
	author = {Tavakol, Mohsen and Dennick, Reg},
	month = jun,
	year = {2011},
	pmid = {28029643},
	pmcid = {PMC4205511},
	pages = {53--55},
}

@inproceedings{cai2022impacts,
  title={Impacts of personal characteristics on user trust in conversational recommender systems},
  author={Cai, Wanling and Jin, Yucheng and Chen, Li},
  booktitle={Proceedings of the 2022 CHI Conference on Human Factors in Computing Systems},
  pages={1--14},
  year={2022}
}

@techreport{gillespie_trust_2023,
	address = {Brisbane, Australia},
	title = {Trust in {Artificial} {Intelligence}: {A} global study},
	shorttitle = {Trust in {Artificial} {Intelligence}},
	url = {https://espace.library.uq.edu.au/view/UQ:00d3c94},
	language = {en},
	urldate = {2023-09-05},
	institution = {The University of Queensland; KPMG Australia},
	author = {Gillespie, Nicole and Lockey, Steven and Curtis, Caitlin and Pool, Javad and {Ali Akbari}},
	month = feb,
	year = {2023},
	doi = {10.14264/00d3c94},
}

@article{schwartz2002maximizing,
  title={Maximizing versus satisficing: happiness is a matter of choice.},
  author={Schwartz, Barry and Ward, Andrew and Monterosso, John and Lyubomirsky, Sonja and White, Katherine and Lehman, Darrin R},
  journal={Journal of personality and social psychology},
  volume={83},
  number={5},
  pages={1178},
  year={2002},
  publisher={American Psychological Association}
}

@inproceedings{schemmer_appropriate_2023,
	address = {New York, NY, USA},
	series = {{IUI} '23},
	title = {Appropriate {Reliance} on {AI} {Advice}: {Conceptualization} and the {Effect} of {Explanations}},
	isbn = {9798400701061},
	shorttitle = {Appropriate {Reliance} on {AI} {Advice}},
	url = {https://dl.acm.org/doi/10.1145/3581641.3584066},
	doi = {10.1145/3581641.3584066},
	urldate = {2023-07-11},
	booktitle = {Proceedings of the 28th {International} {Conference} on {Intelligent} {User} {Interfaces}},
	publisher = {Association for Computing Machinery},
	author = {Schemmer, Max and Kuehl, Niklas and Benz, Carina and Bartos, Andrea and Satzger, Gerhard},
	month = mar,
	year = {2023},
	pages = {410--422},
}

@inproceedings{wang_watch_2023,
	address = {New York, NY, USA},
	series = {{CHI} '23},
	title = {Watch {Out} for {Updates}: {Understanding} the {Effects} of {Model} {Explanation} {Updates} in {AI}-{Assisted} {Decision} {Making}},
	isbn = {978-1-4503-9421-5},
	shorttitle = {Watch {Out} for {Updates}},
	url = {https://dl.acm.org/doi/10.1145/3544548.3581366},
	doi = {10.1145/3544548.3581366},
	urldate = {2023-07-06},
	booktitle = {Proceedings of the 2023 {CHI} {Conference} on {Human} {Factors} in {Computing} {Systems}},
	publisher = {Association for Computing Machinery},
	author = {Wang, Xinru and Yin, Ming},
	month = apr,
	year = {2023},
	pages = {1--19},
}

@article{chong_human_2022,
	title = {Human confidence in artificial intelligence and in themselves: {The} evolution and impact of confidence on adoption of {AI} advice},
	volume = {127},
	issn = {0747-5632},
	shorttitle = {Human confidence in artificial intelligence and in themselves},
	url = {https://www.sciencedirect.com/science/article/pii/S0747563221003411},
	doi = {10.1016/j.chb.2021.107018},
	urldate = {2023-09-03},
	journal = {Computers in Human Behavior},
	author = {Chong, Leah and Zhang, Guanglu and Goucher-Lambert, Kosa and Kotovsky, Kenneth and Cagan, Jonathan},
	month = feb,
	year = {2022},
	pages = {107018},
}

@inproceedings{schoeffer2024explanations,
  title={Explanations, Fairness, and Appropriate Reliance in Human-AI Decision-Making},
  author={Schoeffer, Jakob and De-Arteaga, Maria and Kuehl, Niklas},
  booktitle={Proceedings of the CHI Conference on Human Factors in Computing Systems},
  pages={1--18},
  year={2024}
}

@inproceedings{zhou-etal-2024-relying,
    title = "Relying on the Unreliable: The Impact of Language Models{'} Reluctance to Express Uncertainty",
    author = "Zhou, Kaitlyn  and
      Hwang, Jena  and
      Ren, Xiang  and
      Sap, Maarten",
    editor = "Ku, Lun-Wei  and
      Martins, Andre  and
      Srikumar, Vivek",
    booktitle = "Proceedings of the 62nd Annual Meeting of the Association for Computational Linguistics (Volume 1: Long Papers)",
    month = aug,
    year = "2024",
    address = "Bangkok, Thailand",
    publisher = "Association for Computational Linguistics",
    url = "https://aclanthology.org/2024.acl-long.198",
    pages = "3623--3643",
}

@article{van2005research,
  title={Research on cognitive load theory and its design implications for e-learning},
  author={Van Merrienboer, Jeroen JG and Ayres, Paul},
  journal={Educational Technology Research and Development},
  volume={53},
  number={3},
  pages={5--13},
  year={2005},
  publisher={Springer}
}

@article{KIRSCHNER20021,
title = {Cognitive load theory: implications of cognitive load theory on the design of learning},
journal = {Learning and Instruction},
volume = {12},
number = {1},
pages = {1-10},
year = {2002},
issn = {0959-4752},
doi = {https://doi.org/10.1016/S0959-4752(01)00014-7},
url = {https://www.sciencedirect.com/science/article/pii/S0959475201000147},
author = {Paul A. Kirschner}
}

@article{mousavi1995reducing,
  title={Reducing cognitive load by mixing auditory and visual presentation modes.},
  author={Mousavi, Seyed Yaghoub and Low, Renae and Sweller, John},
  journal={Journal of educational psychology},
  volume={87},
  number={2},
  pages={319},
  year={1995},
  publisher={American Psychological Association}
}

@incollection{sweller2011cognitive,
  title={Cognitive load theory},
  author={Sweller, John},
  booktitle={Psychology of learning and motivation},
  volume={55},
  pages={37--76},
  year={2011},
  publisher={Elsevier}
}

@article{vasconcelos_explanations_2023,
	title = {Explanations {Can} {Reduce} {Overreliance} on {AI} {Systems} {During} {Decision}-{Making}},
	volume = {7},
	url = {https://dl.acm.org/doi/10.1145/3579605},
	doi = {10.1145/3579605},
	number = {CSCW1},
	urldate = {2023-07-13},
	journal = {Proceedings of the ACM on Human-Computer Interaction},
	author = {Vasconcelos, Helena and Jörke, Matthew and Grunde-McLaughlin, Madeleine and Gerstenberg, Tobias and Bernstein, Michael S. and Krishna, Ranjay},
	month = apr,
	year = {2023},
	pages = {129:1--129:38},
}

@inproceedings{Janis1977DecisionMA,
  title={Decision Making: A Psychological Analysis of Conflict, Choice, and Commitment},
author={Janis, Irving L and Mann, Leon},
  year={1977},
  publisher={Free press}
}

@article{deiana2023artificial,
  title={Artificial intelligence and public health: evaluating ChatGPT responses to vaccination myths and misconceptions},
  author={Deiana, Giovanna and Dettori, Marco and Arghittu, Antonella and Azara, Antonio and Gabutti, Giovanni and Castiglia, Paolo},
  journal={Vaccines},
  volume={11},
  number={7},
  pages={1217},
  year={2023},
  publisher={MDPI}
}

@article{dzindolet2003role,
  title={The role of trust in automation reliance},
  author={Dzindolet, Mary T and Peterson, Scott A and Pomranky, Regina A and Pierce, Linda G and Beck, Hall P},
  journal={International journal of human-computer studies},
  volume={58},
  number={6},
  pages={697--718},
  year={2003},
  publisher={Elsevier}
}

@inproceedings{10.1145/3411764.3445522,
author = {Levy, Ariel and Agrawal, Monica and Satyanarayan, Arvind and Sontag, David},
title = {Assessing the Impact of Automated Suggestions on Decision Making: Domain Experts Mediate Model Errors but Take Less Initiative},
year = {2021},
isbn = {9781450380966},
publisher = {Association for Computing Machinery},
address = {New York, NY, USA},
url = {https://doi.org/10.1145/3411764.3445522},
doi = {10.1145/3411764.3445522},
booktitle = {Proceedings of the 2021 CHI Conference on Human Factors in Computing Systems},
articleno = {72},
numpages = {13},
location = {Yokohama,Japan},
series = {CHI '21}
}

@article{morrison_evaluating_2023,
	title = {Evaluating the {Impact} of {Human} {Explanation} {Strategies} on {Human}-{AI} {Visual} {Decision}-{Making}},
	volume = {7},
	issn = {2573-0142},
	url = {https://dl.acm.org/doi/10.1145/3579481},
	doi = {10.1145/3579481},
	language = {en},
	number = {CSCW1},
	urldate = {2023-07-13},
	journal = {Proceedings of the ACM on Human-Computer Interaction},
	author = {Morrison, Katelyn and Shin, Donghoon and Holstein, Kenneth and Perer, Adam},
	month = apr,
	year = {2023},
	pages = {1--37},
}

@article{florenca_food_2021,
	title = {Food {Myths} or {Food} {Facts}? {Study} about {Perceptions} and {Knowledge} in a {Portuguese} {Sample}},
	volume = {10},
	issn = {2304-8158},
	shorttitle = {Food {Myths} or {Food} {Facts}?},
	url = {https://www.ncbi.nlm.nih.gov/pmc/articles/PMC8623929/},
	doi = {10.3390/foods10112746},
	number = {11},
	urldate = {2023-07-13},
	journal = {Foods},
	author = {Florença, Sofia G. and Ferreira, Manuela and Lacerda, Inês and Maia, Aline},
	month = nov,
	year = {2021},
	pmid = {34829026},
	pmcid = {PMC8623929},
	pages = {2746},
}

@article{mann_melbourne_1997,
	title = {The {Melbourne} decision making questionnaire: an instrument for measuring patterns for coping with decisional conflict},
	volume = {10},
	issn = {1099-0771},
	shorttitle = {The {Melbourne} decision making questionnaire},
	url = {https://onlinelibrary.wiley.com/doi/abs/10.1002/%28SICI%291099-0771%28199703%2910%3A1%3C1%3A%3AAID-BDM242%3E3.0.CO%3B2-X},
	doi = {10.1002/(SICI)1099-0771(199703)10:1<1::AID-BDM242>3.0.CO;2-X},
	language = {en},
	number = {1},
	urldate = {2023-07-18},
	journal = {Journal of Behavioral Decision Making},
	author = {Mann, Leon and Burnett, Paul and Radford, Mark and Ford, Steve},
	year = {1997},
	note = {\_eprint: https://onlinelibrary.wiley.com/doi/pdf/10.1002/\%28SICI\%291099-0771\%28199703\%2910\%3A1\%3C1\%3A\%3AAID-BDM242\%3E3.0.CO\%3B2-X},
	pages = {1--19},
}

@article{vodrahalli_humans_2021,
	title = {Do {Humans} {Trust} {Advice} {More} if it {Comes} from {AI}?: {An} {Analysis} of {Human}-{AI} {Interactions}},
	shorttitle = {Do {Humans} {Trust} {Advice} {More} if it {Comes} from {AI}?},
	url = {https://www.semanticscholar.org/reader/8cc80339233348dc6ec2a4a72d66e5b6c2166849},
	doi = {10.1145/3514094.3534150},
	language = {en},
	urldate = {2023-09-03},
	journal = {Proceedings of the 2022 AAAI/ACM Conference on AI, Ethics, and Society},
	author = {Vodrahalli, Kailas and Gerstenberg, Tobias and Zou, James},
	year = {2021},
}

@article{deniz2006relationships,
  title={The relationships among coping with stress, life satisfaction, decision-making styles and decision self-esteem: An investigation with Turkish university students},
  author={Deniz, Mehmet},
  journal={Social Behavior and Personality: an international journal},
  volume={34},
  number={9},
  pages={1161--1170},
  year={2006},
  publisher={Scientific Journal Publishers}
}

@article{kim_nowhere_2022,
	title = {Nowhere else to go: {Help} seeking online and maladaptive decisional styles},
	volume = {128},
	issn = {0747-5632},
	shorttitle = {Nowhere else to go},
	url = {https://www.sciencedirect.com/science/article/pii/S074756322100426X},
	doi = {10.1016/j.chb.2021.107103},
	urldate = {2023-09-03},
	journal = {Computers in Human Behavior},
	author = {Kim, Jisoo and Phillips, James G. and Ogeil, Rowan P.},
	month = mar,
	year = {2022},
	pages = {107103},
}

@article{10.1145/3610219,
author = {Chen, Valerie and Liao, Q. Vera and Wortman Vaughan, Jennifer and Bansal, Gagan},
title = {Understanding the Role of Human Intuition on Reliance in Human-AI Decision-Making with Explanations},
year = {2023},
issue_date = {October 2023},
publisher = {Association for Computing Machinery},
address = {New York, NY, USA},
volume = {7},
number = {CSCW2},
url = {https://doi.org/10.1145/3610219},
doi = {10.1145/3610219},
journal = {Proc. ACM Hum.-Comput. Interact.},
month = {oct},
articleno = {370},
numpages = {32}
}

@article{bubeck2023sparks,
  title={Sparks of artificial general intelligence: Early experiments with gpt-4},
  author={Bubeck, S{\'e}bastien and Chandrasekaran, Varun and Eldan, Ronen and Gehrke, Johannes and Horvitz, Eric and Kamar, Ece and Lee, Peter and Lee, Yin Tat and Li, Yuanzhi and Lundberg, Scott and others},
  journal={arXiv preprint arXiv:2303.12712},
  year={2023}
}

@article{liu2023using,
  title={Using AI-generated suggestions from ChatGPT to optimize clinical decision support},
  author={Liu, Siru and Wright, Aileen P and Patterson, Barron L and Wanderer, Jonathan P and Turer, Robert W and Nelson, Scott D and McCoy, Allison B and Sittig, Dean F and Wright, Adam},
  journal={Journal of the American Medical Informatics Association},
  volume={30},
  number={7},
  pages={1237--1245},
  year={2023},
  publisher={Oxford University Press}
}

@article{hassani2023role,
  title={The role of ChatGPT in data science: how ai-assisted conversational interfaces are revolutionizing the field},
  author={Hassani, Hossein and Silva, Emmanuel Sirmal},
  journal={Big data and cognitive computing},
  volume={7},
  number={2},
  pages={62},
  year={2023},
  publisher={MDPI}
}

@article{10.1093/jla/laae003,
    author = {Dahl, Matthew and Magesh, Varun and Suzgun, Mirac and Ho, Daniel E},
    title = "{Large Legal Fictions: Profiling Legal Hallucinations in Large Language Models}",
    journal = {Journal of Legal Analysis},
    volume = {16},
    number = {1},
    pages = {64-93},
    year = {2024},
    month = {06},
    issn = {2161-7201},
    doi = {10.1093/jla/laae003},
    url = {https://doi.org/10.1093/jla/laae003},
    eprint = {https://academic.oup.com/jla/article-pdf/16/1/64/58336922/laae003.pdf},
}

@inproceedings{bansal2019beyond,
  title={Beyond accuracy: The role of mental models in human-AI team performance},
  author={Bansal, Gagan and Nushi, Besmira and Kamar, Ece and Lasecki, Walter S and Weld, Daniel S and Horvitz, Eric},
  booktitle={Proceedings of the AAAI conference on human computation and crowdsourcing},
  volume={7},
  number={1},
  pages={2--11},
  year={2019}
}

@misc{tao2023auditing,
      title={Auditing and Mitigating Cultural Bias in LLMs}, 
      author={Yan Tao and Olga Viberg and Ryan S. Baker and Rene F. Kizilcec},
      year={2023},
      eprint={2311.14096},
      archivePrefix={arXiv},
      primaryClass={cs.CL}
}

@article{niszczota2023credibility,
  title={The credibility of dietary advice formulated by ChatGPT: robo-diets for people with food allergies},
  author={Niszczota, Pawe{\l} and Rybicka, Iga},
  journal={Nutrition},
  volume={112},
  pages={112076},
  year={2023},
  publisher={Elsevier}
}

@article{MA2023102362,
title = {Are users willing to embrace ChatGPT? Exploring the factors on the acceptance of chatbots from the perspective of AIDUA framework},
journal = {Technology in Society},
volume = {75},
pages = {102362},
year = {2023},
issn = {0160-791X},
doi = {https://doi.org/10.1016/j.techsoc.2023.102362},
url = {https://www.sciencedirect.com/science/article/pii/S0160791X23001677},
author = {Xiaoyue Ma and Yudi Huo}
}

@inproceedings{10.1145/3628454.3629552,
author = {Faruk, Lawal Ibrahim Dutsinma and Rohan, Rohani and Ninrutsirikun, Unhawa and Pal, Debajyoti},
title = {University Students’ Acceptance and Usage of Generative AI (ChatGPT) from a Psycho-Technical Perspective},
year = {2023},
isbn = {9798400708497},
publisher = {Association for Computing Machinery},
address = {New York, NY, USA},
url = {https://doi.org/10.1145/3628454.3629552},
doi = {10.1145/3628454.3629552},
booktitle = {Proceedings of the 13th International Conference on Advances in Information Technology},
articleno = {15},
numpages = {8},
location = { Bangkok,Thailand},
series = {IAIT '23}
}

@inproceedings{yin2019understanding,
  title={Understanding the effect of accuracy on trust in machine learning models},
  author={Yin, Ming and Wortman Vaughan, Jennifer and Wallach, Hanna},
  booktitle={Proceedings of the 2019 chi conference on human factors in computing systems},
  pages={1--12},
  year={2019}
}

@article{feridun2024roles,
author = {Feridun Kaya and Fatih Aydin and Astrid Schepman and Paul Rodway and Okan Yetişensoy and Meva Demir Kaya},
title = {The Roles of Personality Traits, AI Anxiety, and Demographic Factors in Attitudes toward Artificial Intelligence},
journal = {International Journal of Human–Computer Interaction},
volume = {40},
number = {2},
pages = {497-514},
year = {2024},
publisher = {Taylor & Francis},
doi = {10.1080/10447318.2022.2151730},
URL = {https://doi.org/10.1080/10447318.2022.2151730},
eprint = {https://doi.org/10.1080/10447318.2022.2151730}
}

@article{araujo_ai_2020,
	title = {In {AI} we trust? {Perceptions} about automated decision-making by artificial intelligence},
	volume = {35},
	issn = {1435-5655},
	shorttitle = {In {AI} we trust?},
	url = {https://doi.org/10.1007/s00146-019-00931-w},
	doi = {10.1007/s00146-019-00931-w},
	language = {en},
	number = {3},
	urldate = {2024-08-20},
	journal = {AI \& SOCIETY},
	author = {Araujo, Theo and Helberger, Natali and Kruikemeier, Sanne and de Vreese, Claes H.},
	month = sep,
	year = {2020},
	pages = {611--623},
}

@inproceedings{feng-etal-2023-pretraining,
    title = "From Pretraining Data to Language Models to Downstream Tasks: Tracking the Trails of Political Biases Leading to Unfair {NLP} Models",
    author = "Feng, Shangbin  and
      Park, Chan Young  and
      Liu, Yuhan  and
      Tsvetkov, Yulia",
    editor = "Rogers, Anna  and
      Boyd-Graber, Jordan  and
      Okazaki, Naoaki",
    booktitle = "Proceedings of the 61st Annual Meeting of the Association for Computational Linguistics (Volume 1: Long Papers)",
    month = jul,
    year = "2023",
    address = "Toronto, Canada",
    publisher = "Association for Computational Linguistics",
    url = "https://aclanthology.org/2023.acl-long.656",
    doi = "10.18653/v1/2023.acl-long.656",
    pages = "11737--11762",
}

@inproceedings{bo2025to,
author = {Bo, Jessica Y and Wan, Sophia and Anderson, Ashton},
title = {To Rely or Not to Rely? Evaluating Interventions for Appropriate Reliance on Large Language Models},
year = {2025},
isbn = {9798400713941},
publisher = {Association for Computing Machinery},
address = {New York, NY, USA},
url = {https://doi.org/10.1145/3706598.3714097},
doi = {10.1145/3706598.3714097},
booktitle = {Proceedings of the 2025 CHI Conference on Human Factors in Computing Systems},
articleno = {905},
numpages = {23},
location = {
},
series = {CHI '25}
}

@inproceedings{caraban2019nudge,
author = {Caraban, Ana and Karapanos, Evangelos and Gon\c{c}alves, Daniel and Campos, Pedro},
title = {23 Ways to Nudge: A Review of Technology-Mediated Nudging in Human-Computer Interaction},
year = {2019},
isbn = {9781450359702},
publisher = {Association for Computing Machinery},
address = {New York, NY, USA},
url = {https://doi.org/10.1145/3290605.3300733},
doi = {10.1145/3290605.3300733},
booktitle = {Proceedings of the 2019 CHI Conference on Human Factors in Computing Systems},
pages = {1–15},
numpages = {15},
location = {Glasgow, Scotland Uk},
series = {CHI '19}
}

This paper has no relation to authors’ prior publications.
\appendix
\newpage
\section{Nutrition Statements}

\begin{table}[!h]
\begin{tabularx}{\textwidth}{|c|X|c|c|}
\hline
Code & Statement & Nature & Presented AI Decision \\ 
\hline
S1 & Drinking water during meals, contributes to weight gain. & Myth & Myth\\
S2 & The digestion process begins in the mouth. & Fact  & Fact\\
S3 & Fruit should be eaten before meals. & Myth & Myth\\
S4 & Egg consumption increases blood cholesterol. & Myth & Myth\\
S5 & Drinking milk is bad for health. & Myth& Myth\\
S6 & Eating carbohydrates at night leads to an increase in weight gain. & Myth & Fact\\
S7 & Fat is important to the human body. & Fact  & Fact\\
S8 & Fruit should be eaten after meals. & Myth& Myth \\
S9 & Fiber intake is important for normal bowel function. & Fact &Fact\\
S10 & Gluten-free foods are better for health and should, there-fore, be adopted by all. & Myth & Myth\\
S11 & Cheese consumption is bad for memory. & Myth & Myth\\
S12 & Coconut oil is healthier than olive oil. & Myth & Myth\\
S13 & Lactose-free foods are better for health and should, there-fore, be adopted by all. & Myth & Myth\\
S14 & Children have different nutritional needs than those for adults. & Fact  & Fact\\
S15 & Fruits and vegetables do not contribute to weight gain. & Myth & Myth\\
S16 & Normal potatoes are more caloric than sweet potatoes. & Myth& Fact\\
S17 & Diet should be adapted to a person’s blood group. & Myth & Fact\\
S18 & Not having a balanced and varied diet can lead to the development of multiple diseases. & Fact & Myth\\
S19 & The alkaline diet allows balancing the acidity in the blood. & Myth & Myth\\
S20 & Drinking, while fasting, a glass of water with lemon helps in weight loss. & Myth&Fact\\
S21 & Inadequate eating habits are the third risk factor for the loss of years of healthy life. & Fact  & Fact\\
S22 & Ingesting high amounts of protein helps in the faster formation of muscles. & Myth & Fact\\
S23 & Pregnant women should be eating for two. & Myth & Fact\\
S24 & Cold water should not be drunk. & Myth & Myth\\
S25 & The day should always start with breakfast. & Fact  & Myth\\
S26 & Water is essential to the normal function of all organs. & Fact  & Fact\\
S27 & Soy milk is healthier than cow’s milk. & Myth & Myth\\
S28 & Orange should not be eaten at the same time as milk or yogurt. & Myth & Myth\\
S29 & Dairy products should be consumed in between two and three portions per day. & Fact  & Myth\\
S30 & All food additives (E’s) are harmful to health. & Myth &Myth\\

\hline
\end{tabularx}
\caption{Nutrition statements and presented AI's decisions used in the study's experiment for participants. These statements are adopted from \citet{florenca_food_2021}. }
\Description{This image is a table that displays nutrition statements and the corresponding AI's decision on whether the statement is a myth or a fact. The table has three columns: Code, Statement, and Nature/Presented AI Decision.The statements cover a range of topics related to nutrition, such as the effects of consuming certain foods or beverages, dietary recommendations, and general health claims. The "Nature" column indicates whether the statement is a myth or a fact, and the "Presented AI Decision" column shows whether the AI deemed the statement to be a myth or a fact.The table covers 30 different nutrition statements, with the AI classifying 12 as facts and 18 as myths. Some examples of the statements include "Drinking water during meals, contributes to weight gain" (classified as a myth), "The digestion process begins in the mouth" (classified as a fact), and "Gluten-free foods are better for health and should, there-fore, be adopted by all" (classified as a myth).Overall, this table provides a comprehensive overview of the AI's assessment of various nutrition-related claims, which could be useful for understanding the system's knowledge and decision-making process in this domain.}
\end{table}

\newpage
\section{The Melbourne Decision-Making Questionnaire}

\begin{table}[!h]
\caption{The Melbourne Decision-Making Questionnaire by \citet{mann_melbourne_1997}, which we used in our study to assess participants' decision-making patterns. All items could be answered on a 3-point scale: ``Not true for me'', ``Sometimes true for me'', or ``True for me''.}\label{table:questionnaire}
\begin{tabularx}{\textwidth}{p{2cm}p{12cm}}
\toprule
\textbf{Vigilance} & Item \\ 

1 & I like to consider all of the alternatives.\\
2 & I try to find out the disadvantages of all alternatives.\\
3 & I consider how best to carry out a decision.\\
4 & When making decisions I like to collect a lot of information.\\
5 & I try to be clear about my objectives before choosing.\\
6 & I take a lot of care before choosing.\\
\hline 
\textbf{Buckpassing} & \\
 
7 & I avoid making decisions.\\
8 & I do not make decisions unless I really have to.\\
9 & I prefer to leave decisions to others.\\
10 & I do not like to take responsibility for making decisions.\\
11 & If a decision can be made by me or another person I let the other person make it.\\
12 & I prefer that people who are better informed decide for me.\\
\hline 
\textbf{Hypervigilance} & \\

13 & Whenever I face a difficult decision I feel pessimistic about finding a good solution.\\
14 & I feel as if I am under tremendous time pressure when making decisions.\\
15 & The possibility that some small thing might go wrong causes me to swing abruptly in my preference.\\
16 & I cannot think straight if I have to make a decision in a hurry.\\
17 & After a decision is made I spend a lot of time convincing myself it was correct.\\ 
\bottomrule
\end{tabularx}
\Description{The Melbourne Decision-Making Questionnaire by Mann et al. [44], used to assess participants' decision-making patterns. Items are answered on a 3-point scale: 'Not true for me,' 'Sometimes true for me,' or 'True for me.'  
- Vigilance (Items 1–6): Statements focus on considering alternatives, analyzing disadvantages, collecting information, and being deliberate and clear in decision-making.  
- Buckpassing (Items 7–12): Statements highlight avoiding or deferring decisions, preferring others to decide, and avoiding responsibility.  
- Hypervigilance (Items 13–17): Statements express feeling pressured, pessimistic, or anxious about decisions, and struggling with rushed or uncertain choices.}
\end{table}

\subsection{Study Design}
To ensure that participants were aware of any inaccurate information we presented to them during the study, we included at the end of the study a page about their performance in the nutrition evaluation task. Specifically, we included statements where they submitted inaccurate answers and the correct answers for them as well as statements where AI suggestions were not accurate.
\begin{figure}[h]
    \centering
    \includegraphics[width=0.5\linewidth]{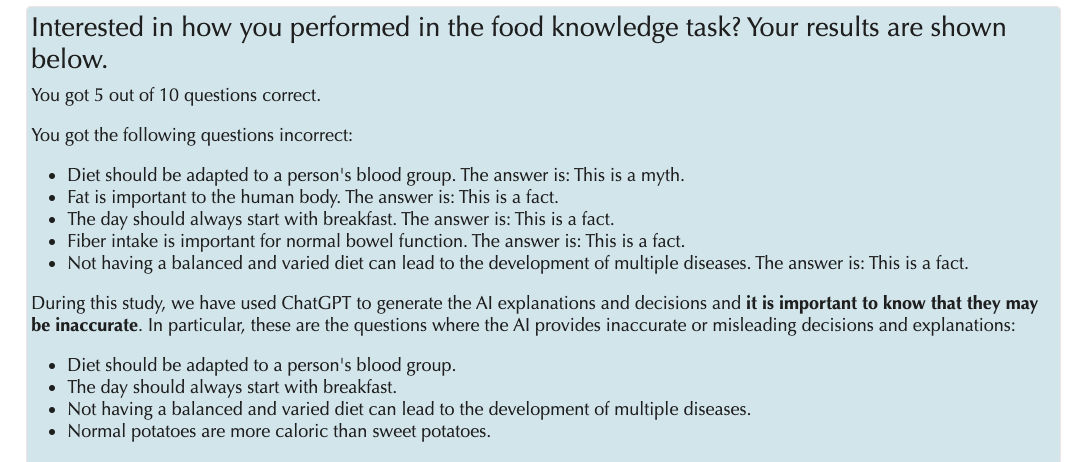}
    \caption{Example of result page in the study presented to participants: we included an example of our result page regarding how we debriefed participants about their performance in the nutrition evaluation task. Specifically, we included the statements where they submitted inaccurate answers and statements where AI suggestions were inaccurate. }
    \Description{An image showing an example of a result page presented to participants in a study. The page provides feedback on participants' performance in a food knowledge task, showing that they answered 5 out of 10 questions correctly. It lists the questions they got wrong and the correct answers, such as "Diet should be adapted to a person's blood group. The answer is: This is a myth," and "The day should always start with breakfast. The answer is: This is a fact." Additionally, the page explains that ChatGPT was used to generate AI-based explanations and decisions, noting that some may be inaccurate. Specific questions where AI suggestions were inaccurate are highlighted, including topics like dietary adaptation to blood groups and the caloric comparison between normal and sweet potatoes.}
\end{figure}

\subsection{Factor Analysis for Decision-making Patterns}
\begin{figure}[h]
    \centering
    \includegraphics[width=0.9\linewidth]{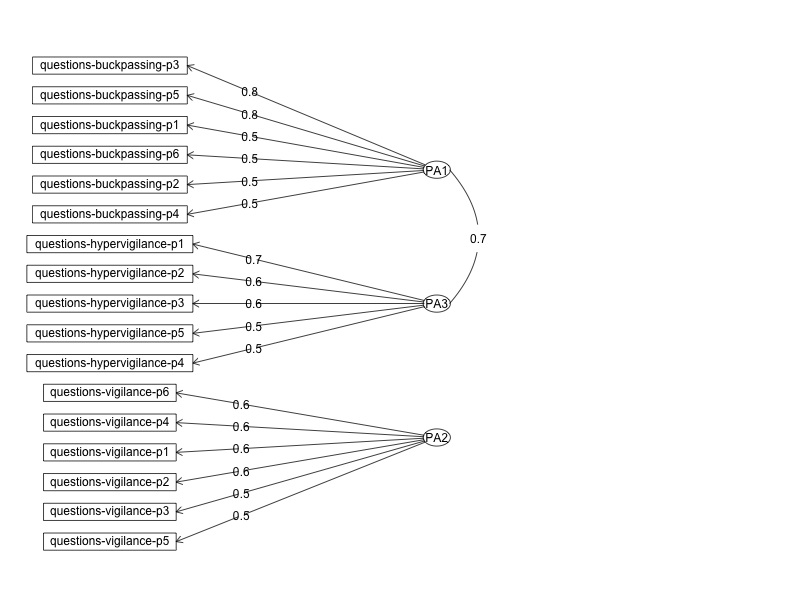}
    \caption{ Factor loading diagram confirming the three decision-making patterns based on the Melbourne Decision Making Questionnaire. Items associated with buckpassing load strongly on PA1, items related to vigilance load on PA2, and items linked to hypervigilance load on PA3, supporting the theoretical structure of the questionnaire. The inter-factor correlation between PA1 and PA3 is shown with a value of 0.7. Loadings represent the strength of the relationship between items and factors, with higher values indicating stronger associations.}
    \Description{A factor loading diagram illustrating the three decision-making patterns based on the Melbourne Decision-Making Questionnaire. The diagram includes three latent factors: PA1 (Buckpassing), PA2 (Vigilance), and PA3 (Hypervigilance). Items related to buckpassing (e.g., questions-buckpassing-p1 through p6) have strong loadings on PA1, items related to vigilance (e.g., questions-vigilance-p1 through p6) load on PA2, and items related to hypervigilance (e.g., questions-hypervigilance-p1 through p5) load on PA3. The inter-factor correlation between PA1 and PA3 is depicted with a value of 0.7. Each loading value (ranging from 0.5 to 0.8) represents the strength of the relationship between individual items and their respective factors, with higher values indicating stronger associations.}
    \label{fig:factor-analysis}
\end{figure}

\newpage
\section{Demographics Information of Participants}
\begin{figure}[!h]
    \centering
    \includegraphics[width=0.75\linewidth]{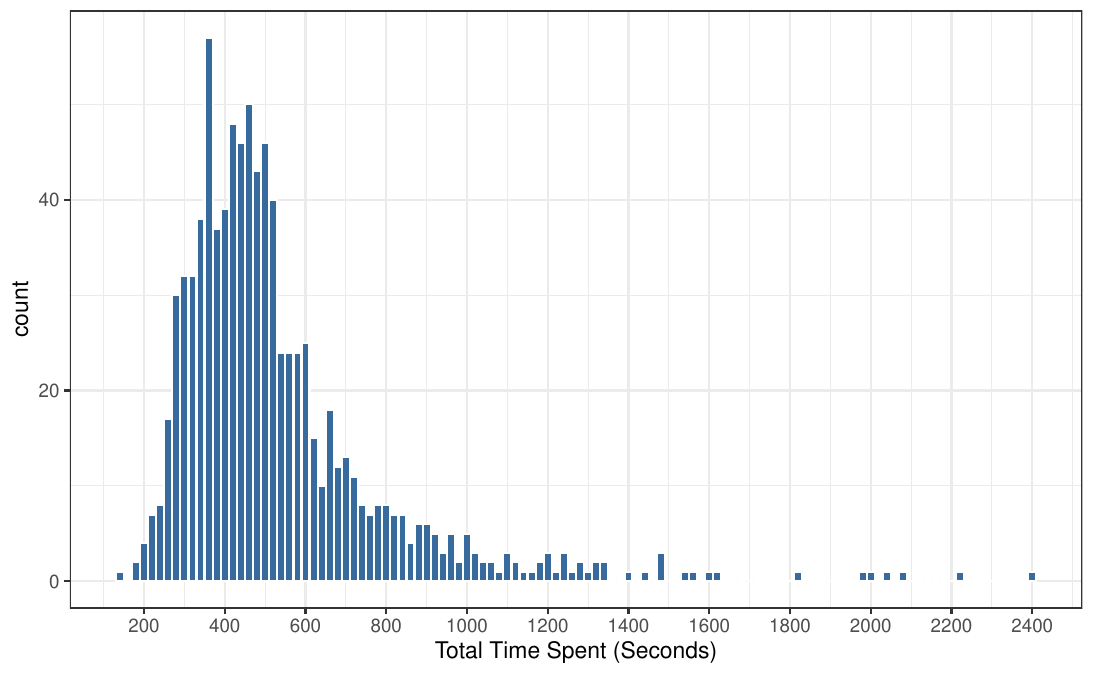}
    \caption{Overall distribution of total time spent in the study: we selected time thresholds (240 seconds and 1200 seconds) where the number of participants significantly dropped and then filtered out participants based on these time threshold.}
    \label{fig:distribution}
\end{figure}

\begin{figure}[!h]
    \centering
    \includegraphics[width=0.6\linewidth]{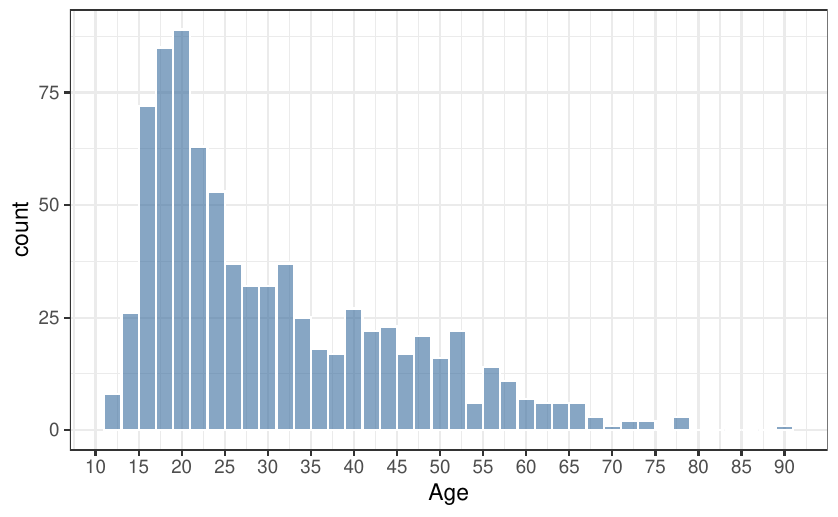}
    \caption{Age distribution of 810 participants.}
    \label{fig:enter-label}
    \Description{A histogram showing the age distribution of 810 participants. The x-axis represents age, ranging from 12 to 90 years, while the y-axis represents the count of participants. The distribution is skewed towards younger ages, with the highest count in the 15–25 age range. The number of participants decreases steadily with age, showing a sharp decline after 30 and tapering off significantly beyond 60. Only a few participants are above 70 years old.}
\end{figure}

\begin{table*}
\centering
\caption{Sample sizes as well as averages and standard deviations for each decision-making measure across different demographics. The maximum for vigilance and buckpassing is 12, whereas the maximum for hypervigilance is 10.}
\label{dm_stats_table}

\begin{tabular}{lrlll}
\hline
 & N & Vigilance & Hypervigilance & Buckpassing\\

\multicolumn{5}{l}{\textbf{Overall}}\\
\hline
\hspace{1em}Overall & 810 & 9.41 $\pm$ 2.31 & 4.25 $\pm$ 2.61 & 4.56 $\pm$ 3.03\\

\multicolumn{5}{l}{\textbf{Education}}\\

\hspace{1em}<=High School Education & 231 & 9.39 $\pm$ 2.32 & 4.94 $\pm$ 2.55 & 5.25 $\pm$ 3.12\\

\hspace{1em}College Education & 403 & 9.34 $\pm$ 2.34 & 4.12 $\pm$ 2.64 & 4.49 $\pm$ 3\\

\hspace{1em}Graduate Education & 176 & 9.59 $\pm$ 2.2 & 3.64 $\pm$ 2.45 & 3.82 $\pm$ 2.78\\
\hline
\multicolumn{5}{l}{\textbf{Gender}}\\

\hspace{1em} Female & 409 & 9.43 $\pm$ 2.31 & 4.52 $\pm$ 2.63 & 4.69 $\pm$ 3.15 \\
   \hspace{1em}     Male & 363 & 9.36 $\pm$ 2.32 & 3.83 $\pm$ 2.47 & 4.33 $\pm$ 2.84 \\
  \hspace{1em}  Other & 38 & 9.66 $\pm$ 2.13 & 5.5 $\pm$ 2.94 & 5.47 $\pm$ 3.27 \\
\hline
\multicolumn{5}{l}{\textbf{Age Group}}\\

\hspace{1em}Under 18 & 155 & 9.30 $\pm$ 2.35 & 5.32 $\pm$ 2.43 & 5.59  $\pm$ 3.09\\
 \hspace{1em}18--24 & 215 & 9.69 $\pm$ 2.22 & 4.64 $\pm$ 2.52 & 5.01 $\pm$ 2.97 \\
 \hspace{1em}       25--34 & 174 & 9.32 $\pm$ 2.51 & 4.37 $\pm$ 2.6  & 4.87 $\pm$ 3.03 \\
  \hspace{1em}      35--44 & 112 & 9.53 $\pm$ 2.24 & 3.49 $\pm$ 2.48 & 3.7 $\pm$ 2.78 \\
  \hspace{1em}      45--54 & 88  & 9.03 $\pm$ 2.19 & 3.05 $\pm$ 2.45 & 3.03 $\pm$ 2.58 \\
   \hspace{1em}     Over 55 & 66  & 9.26 $\pm$ 2.14 & 3.14 $\pm$ 2.42 & 3.39 $\pm$ 2.6 \\

\hline






\hline
\end{tabular}
\end{table*}

\vspace{5mm} 

\newpage

\section{Correlation Matrix for All Variables}
\begin{table}[ht]
\centering
\caption{Correlation matrix of demographic variables. *** significance at the .001 level; ** significance at the .01 level; * significance at the .05 level; . significance at the 0.1 level.}
\begin{tabular}{lcccccccc}
\hline
 & B & V & H & A & HE & CE & GE & DKE \\ 
\hline
Buckpassing (B) & 1 &  &  &  &  &  &  &  \\ 
Vigilance (V) & -0.05 & 1 &  &  &  &  &  &  \\ 
Hypervigilance (H) & 0.59*** & 0.05 & 1 &  &  &  &  &  \\ 
Age (A) & -0.27*** & -0.05 & -0.28*** & 1 &  &  &  &  \\ 
Less than high school education (HE) & 0.14*** & -0.00 & 0.17*** & -0.35*** & 1 &  &  &  \\ 
Pursuing or Have obtained college education (CE) & -0.02 & -0.03 & -0.05 & 0.11** & -0.63*** & 1 &  &  \\ 
Pursuing or Have obtained graduate education (GE) & -0.13*** & 0.04 & -0.12*** & 0.25*** & -0.33*** & -0.52*** & 1 &  \\ 
Domain Knowledge (Expert) (DKE) & -0.20*** & 0.07* & -0.20*** & 0.22*** & -0.15*** & 0.00 & 0.16*** & 1 \\ 
\hline
\end{tabular}
\label{correlation_matrix}
\end{table}


\begin{figure}
    \centering
    \includegraphics[width=0.75\linewidth]{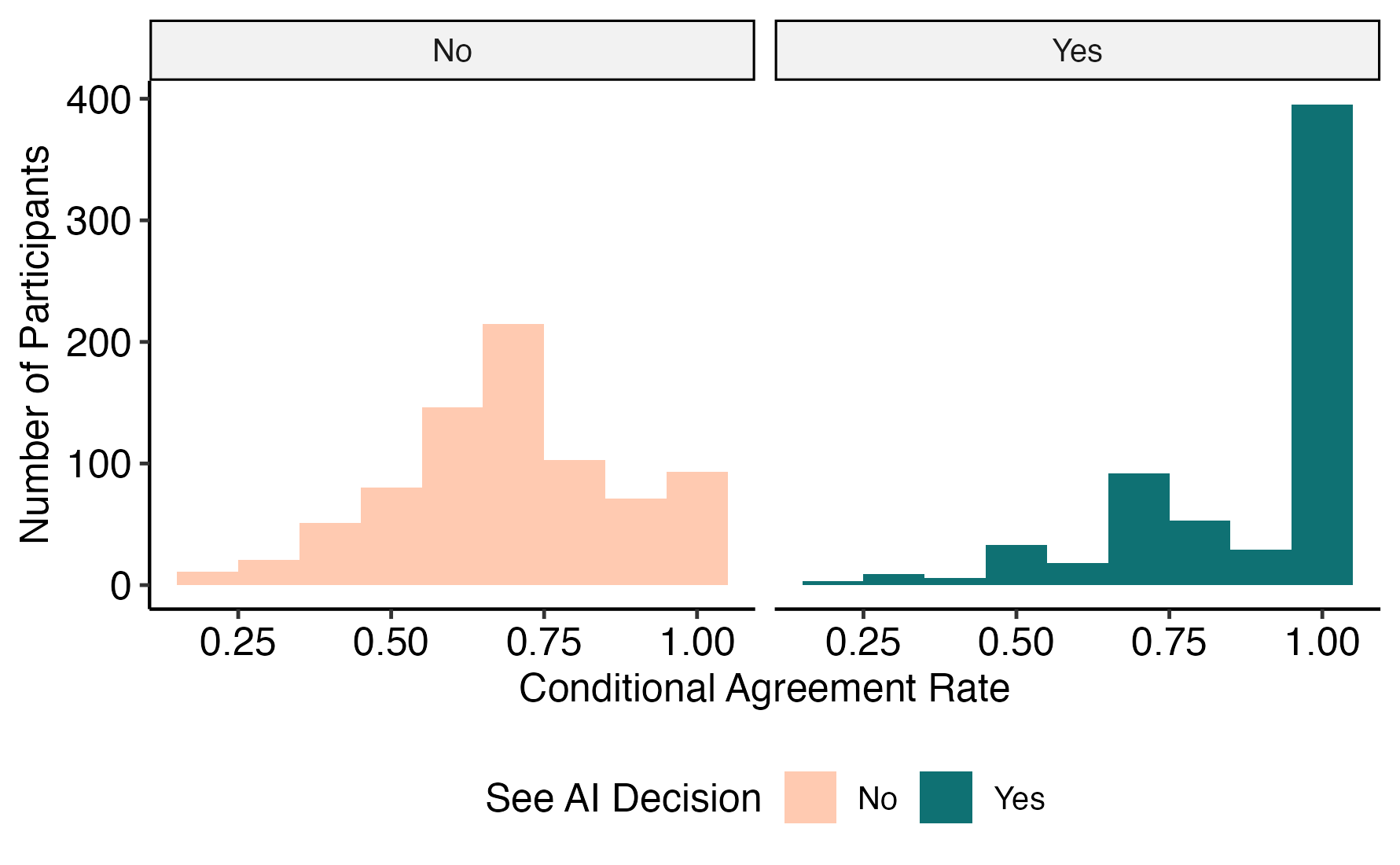}
    \caption{Distribution of Conditional Agreement with AI Decision Rate: conditional agreement rate is participants' agreement rate conditioned on whether they chose to see AI decisions. There are 395 out of 810 participants (49\%) who agreed with AI decisions when they chose to see them.}
    \label{fig:agreement_rate}
\end{figure}

\begin{figure}
    \centering  \includegraphics[width=0.9\linewidth]{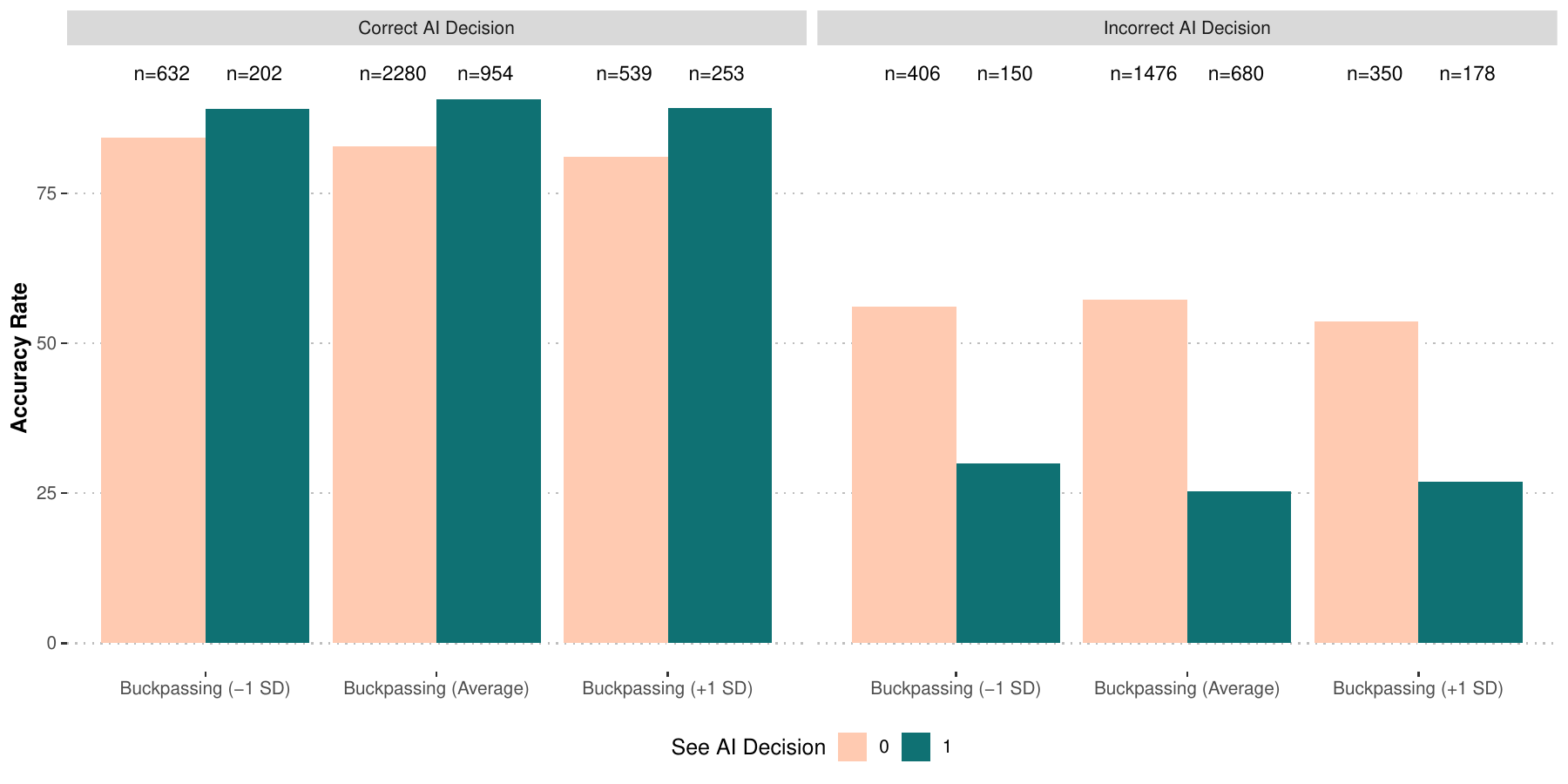}
    \caption{Overall performance breakdown among participants with different levels of buckpassing: participants who chose to see AI decisions showed higher accuracy when the AI’s decisions were correct and lower accuracy when the AI’s decisions were incorrect, across all levels of buckpassing. When exposed to incorrect AI decisions, participants with high buckpassing scores (45\%) showed lower accuracy than those with low buckpassing scores (49\%). When exposed to correct AI decisions, participants with high buckpassing scores (84\%) performed similarly to those with low buckpassing scores (85\%). The n indicates the number of samples in each group and condition.}
    \label{fig:accuracy_breakdown}
\end{figure}

\newpage
\section{Additional Regression Analyses}
\begin{table}[ht]
\caption{Mixed-effects logistic regression results predicting See\_AI\_Explanations: no statistically significant effect of decision-making patterns is observed.}
\centering
\begin{tabular}{lcccc}
\hline
\textbf{Fixed Effects} & \textbf{Estimate} & \textbf{Std. Error} & \textbf{z value} & \textbf{p-value} \\
\hline
(Intercept)      & -1.9223  & 0.32228 & -5.965 & $<$ 0.001 *** \\
buckpassing      & 0.48502  & 0.31146 & 1.557  & 0.119 \\
vigilance        & -0.02133 & 0.32836 & -0.065 & 0.948 \\
hypervigilance   & -0.18182 & 0.29880 & -0.608 & 0.543 \\
\hline
\end{tabular}

\label{tab:regression_results_seeAIexp}
\end{table}

\begin{table}[ht]
\centering
\caption{Mixed-effects logistic regression result predicting whether participants correctly evaluate whether each nutrition statement is a fact or a myth: result indicates no effects of decision-making patterns. However, participants who chose to see AI decisions and see AI explanations were more likely to have incorrect responses.}
    \begin{tabular}{l c c}
        \hline
        \textbf{Predictors} & \textbf{Odds Ratios} & \textbf{CI} \\
        \hline
        (Intercept) & 5.56$^{***}$ & 2.86 -- 10.79 \\
         Buckpassing & 1.00 & 0.98 -- 1.03 \\
        Hypervigilance & 0.98 & 0.95 -- 1.01 \\
 
        Vigilance & 0.99 & 0.96 -- 1.02 \\
        Age & 1.00 & 0.99 -- 1.00 \\
        Domain Knowledge (Average) & 0.88 & 0.77 -- 1.02 \\
        Domain Knowledge (Below Average) & 0.80$^{*}$ & 0.65 -- 0.97 \\
        Pursuing or Have obtained college education & 1.10 & 0.94 -- 1.29 \\
        Pursuing or Have obtained graduate education & 1.20 & 0.99 -- 1.46 \\
     
        \hline
        N Participant & 810 & \\
        N statement & 30 & \\
        \hline
        Observations & 8100 & \\
        Marginal R$^2$ / Conditional R$^2$ & 0.003 / 0.432 & \\
        \hline
    \end{tabular}
\end{table}

\end{document}